\documentclass[journal]{IEEEtran}

\usepackage{subfigure}
\usepackage{colortbl}
\usepackage{bm}

\usepackage{graphicx}  
\usepackage{url}       

\usepackage{amsmath}    
\allowdisplaybreaks[4]
\usepackage{cite}

\usepackage{amsfonts,amssymb}

\usepackage{stfloats}

\usepackage{cases}
\usepackage{algorithm}
\usepackage{multirow}
\usepackage{algorithmic}
\usepackage{epstopdf}
\usepackage{color}
\usepackage{bigints}
\usepackage{mathtools}
\usepackage{relsize}
\usepackage[square, comma, sort&compress, numbers]{natbib}

\makeatletter
\newcommand{\vvast}{\bBigg@{3.0}}
\newcommand{\vast}{\bBigg@{4}}
\newcommand{\Vast}{\bBigg@{4.5}}
\newcommand{\VVast}{\bBigg@{5}}
\newcommand{\VVVast}{\bBigg@{5.5}}
\newcommand{\VVVVast}{\bBigg@{6}}

\makeatother

\newtheorem{theorem}{{Theorem}}

\newcommand{\ls}[1]
{\dimen0=\fontdimen6\the\font
	\lineskip=#1\dimen0
	\advance\lineskip.5\fontdimen5\the\font
	\advance\lineskip-\dimen0
	\lineskiplimit=.9\lineskip
	\baselineskip=\lineskip
	\advance\baselineskip\dimen0
	\normallineskip\lineskip
	\normallineskiplimit\lineskiplimit
	\normalbaselineskip\baselineskip
	\ignorespaces
}

\ifCLASSINFOpdf
\else
\fi
\begin{document}
	\pagenumbering{gobble}
	

	\title{\ls{1.0}{Performance Boundary Analyses for Statistical Multi-QoS Framework Over 6G SAGINs}}

\author{\IEEEauthorblockN{Jingqing Wang,~\IEEEmembership{Member,~IEEE}, Wenchi Cheng,~\IEEEmembership{Senior Member,~IEEE}, and Wei Zhang,~\IEEEmembership{Fellow,~IEEE}} \\[0.2cm]
	
	\thanks{
		This work was supported in part by National Key R\&D Program of China under Grant 2021YFC3002100.}
	\thanks{Jingqing Wang and Wenchi Cheng are with the State Key Laboratory of Integrated Services Networks, Xidian University, Xi’an, China (e-mails: wangjingqing00@gmail.com; wccheng@xidian.edu.cn).}
	\thanks{Wei Zhang is with the School of Electrical Engineering and Telecommunications, The University of New South Wales, Sydney, Australia (e-mail: w.zhang@unsw.edu.au).}
}

\maketitle

\begin{abstract}
To enable the cost-effective universal access and the enhancement of current communication services, the space-air-ground integrated networks (SAGINs) have recently been developed due to its exceptional 3D coverage and the ability to guarantee rigorous and multidimensional demands for quality-of-service (QoS) provisioning, including delay and reliability across vast distances.
The integration of spatial, aerial, and terrestrial dimensions is thus regarded as a critical facilitator for accommodating massive Ultra-Reliable Low-Latency Communications (mURLLC) applications.
In response to the complex, heterogeneous, and dynamic serving scenarios and stringent performance expectations for 6G SAGINs, it is crucial to undertake modeling, assurance, and analysis of the key technologies, aligned with the diverse demands for QoS provisioning in the non-asymptotic regime, i.e., when implementing finite blocklength coding (FBC) as a new dimension for error-rate bounded QoS metric.
However, how to design new statistical QoS-driven performance modeling approaches that accurately delineate the complex and dynamic behaviors of networks, particularly in terms of constraining both delay and error rate, persists as a significant challenge for implementing mURLLC within 6G SAGINs in the finite blocklength regime.
To overcome these difficulties, in this paper we propose to develop a set of analytical modeling frameworks for 6G SAGIN in supporting statistical delay and error-rate bounded QoS in the finite blocklength regime.
First we establish the SAGIN system architecture model. 
Second, the aggregate interference and decoding error probability functions are modeled and examined through using Laplace transform. 
Third, we introduce modeling techniques aimed at defining the$\epsilon$-effective capacity function as a crucial metric for facilitating statistical QoS standards with respect to delay and error-rate. 
To validate the effectiveness of the developed performance modeling schemes, we have executed a series of simulations over SAGINs.
\end{abstract}


\section{Introduction}\label{sec:intro}

The emergence of 6G wireless networks marks a significant leap beyond terrestrial, enabling a comprehensive communication framework that integrates the spatial, aerial, and terrestrial dimensions of network architecture~\cite{9003618}. 
The vision of future 6G 3D radio environments can be realized by deploying satellites and \textit{unmanned aerial vehicles} (UAVs)~\cite{9121338} as additional platforms/tiers, while effectively mitigating aggregate interference within controllable bounds.
Such space-air-ground integrated network (SAGIN) architectures~\cite{8368236} offer unique advantages to significantly improve various performances, including seamless, high-quality connectivities and unparalleled coverage expanses while guaranteeing an unprecedented level of ultra reliability and low delay, etc.
Towards this end, the emergence and subsequent recognition of SAGINs as a robust and promising solution for supporting the delay/error-sensitive services of mURLLC represents a significant advancement~\cite{BANAFAA2023245,YEH202382}.

However, the innovative SAGINs face many new challenges. 
In response to the complex, heterogeneous, and dynamic serving scenarios and stringent performance expectations confronting 6G SAGINs, it is crucial to undertake modeling, assurance, and analysis of the key technologies, aligned with the rigorous and multidimensional demands for quality-of-service (QoS) provisioning, including delay, reliability, etc.
Accordingly, the theory of \textit{statistical delay-bounded QoS provisioning}~\cite{10177877,10436889,10609803,10494937}, emerges as a potent methodology for encapsulating the assurance of delay violation probability, 
offering a novel paradigm for addressing the challenge in terms of delay-sensitive real-time wireless services.
However, the rapid advancements in 6G era, combined with the exponential increase in delay-sensitive real-time services, impose increasingly stringent and diverse QoS requirements.
Thus, it is necessary to design novel diverse QoS provisioning strategies, which are capable of navigating the complexities of 6G network environments, thereby ensuring robust, efficient, and reliable communication for mURLLC services.

To support the advanced and diverse QoS provisioning with envisioned SAGINs in 3D radio environments, it is of paramount significance to model and analyze the reliability QoS requirement in the non-asymptotic regime, i.e., when considering the finite blocklength coding (FBC)~\cite{Yp2011,Yp2014,elwekeil2022power,9635675}.
Previous studies have developed the FBC techniques to support \textit{small-packet communications} with non-vanishing decoding error probability.
The authors of~\cite{9558857} have presented a comprehensive survey about the current state-of-the-art on UAV-enabled URLLC networks by illustrating the main features and implementation challenges in the finite blocklength regime.
The authors of~\cite{9686609} have invested the age-oriented hybrid automatic repeat request (HARQ) protocol design over SAGINs using FBC.
The authors of~\cite{10353011} have characterized the queuing latency and transmission reliability of uplink and downlink SAGINs to improve URLLC services.

Although the integration of space, air, and ground dimensions within current mobile wireless network architecture endeavors to fulfill the requisite QoS performance standards, the saturation of the ultra-high frequency (UHF) spectrum alongside the escalating demands for wireless traffic has precipitated the necessity for incorporating UAVs that operate on higher-frequency, such as millimeter waves (mmWave)~\cite{9598918,8876665}. 
The recourse of mmWave communications underscores a critical shift towards optimizing network capacity and reliability, addressing the bandwidth necessities for mURLLC applications while ensuring high data-rate demands anticipated within the 6G era. 

Nonetheless, while offering substantial benefits for the bandwidth and data rate requirements, the deployment of mmWave systems introduces specific challenges, including substantial pathloss and the diminished penetration capabilities at mmWave band, which could undermine the effective deployment of SAGINs. 
In response, researchers have examined both the technical potential and the challenges for mmWave-based UAV networks~\cite{9768113}. 
Concurrently, the authors of~\cite{9640476} have formulated a problem to jointly optimize the UAV positioning, user clustering, and hybrid analog-digital beamforming for the maximization of user achievable sum rate.
Further investigations, such as those presented in~\cite{10181128}, develops a 3D irregular shaped geometry-based stochastic model for mmWave UAV based massive multiple-input multiple-output (MIMO) vehicle-to-vehicle channels. 
Moreover, the authors of~\cite{9315157} have explored the dual-mode UAV-assisted service delivery in mmWave band for achieving a trade-off between QoS and energy consumption.

Upon integrating spatial, aerial, and terrestrial dimensions of network architectures, the exploration of SAGIN framework is still at a preliminary phase.
The emergence of such an advanced, integrated network architecture requires to be configured/characterized by diverse QoS measuring/controlling mechanisms and new methodologies for the formulation and assessment of system performance indicators. 
This lack of diversity in current network performance modeling/controlling mechanisms leads to restrictions in the expansion of system capacity and diminishes the adaptability of SAGINs, ultimately obstructing essential advancements in efficiency.
Consequently, designing novel statistical QoS-driven performance modeling approaches that accurately delineate the complex and dynamic behaviors of networks, particularly constraining delay and error rate, persists as a significant challenge for implementing mURLLC over 6G SAGINs in the context of the finite blocklength regime.

To effectively overcome the above-mentioned challenges, in this paper we propose to develop the statistical delay and error-rate bounded QoS-driven performance modeling schemes over SAGINs in the finite blocklength regime.
Specifically, our proposed SAGIN architecture consists of the GBSs, UAVs, and satellites.
Subsequently, we construct a suite of analytical modeling frameworks designed to encapsulate the $\epsilon$-effective capacity, facilitating the provision of statistical delay and error-rate bounded QoS for FBC-based SAGINs. 
First, we build wireless communication models in the finite blocklength domain. 
Second, the aggregate interference, decoding error probability, and outage probability functions are modeled and examined through using the Laplace transform. 
Furthermore, we introduce modeling methodologies aimed at defining and analyzing the $\epsilon$-effective capacity through deriving the outage capacity to ensure rigorous and diverse QoS through FBC.
Finally, we perform a series of simulations to validate and assess the developed performance modeling schemes for mURLLC over SAGINs.

The rest of this paper is organized as follows. Section~\ref{sec:sys} builds the system architecture models over SAGINs and the channel coding model through applying FBC.
Section~\ref{sec:DE} characterizes the aggregate interference, decoding error probability, and outage probability, respectively.
Section~\ref{sec:EC} proposes to model the $\epsilon$-effective capacity through using FBC. 
Section~\ref{sec:results} conducts a set of simulations to analyze and assess the system performance.  The paper concludes with Section~\ref{sec:conclusion}.

	\section{The System Architecture Models over SAGINs}\label{sec:sys}

\begin{figure}[!t]
	\centering
	\includegraphics[scale=0.38]{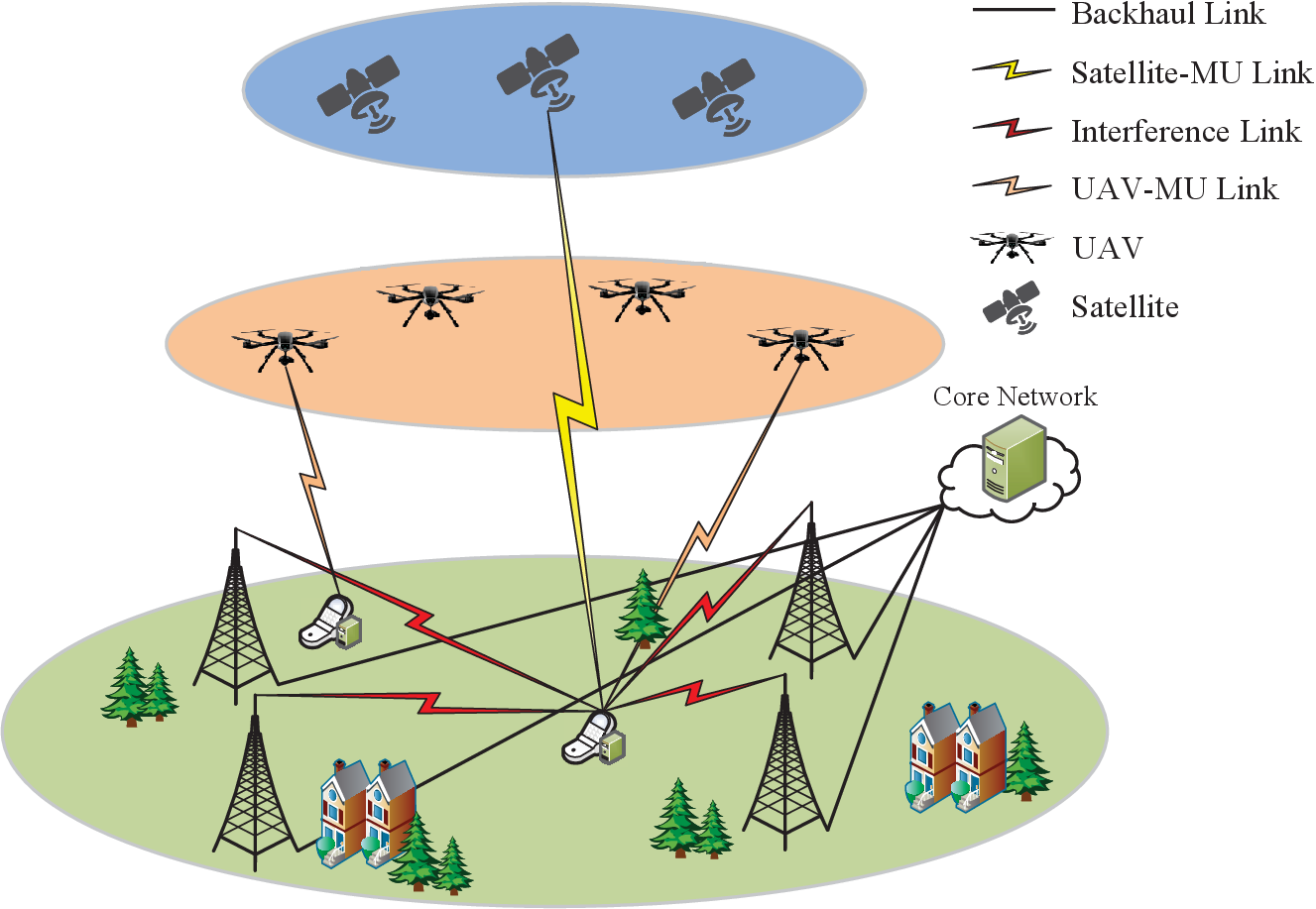}
	\caption{The system architecture model for the developed 6G SAGINs for guaranteeing statistical QoS provisioning in terms of delay and error-rate.}
	\label{fig:1}
\end{figure}

{Fig.~\ref{fig:1}} illustrates the network architecture model for the developed 6G SAGINs for guaranteeing statistical QoS provisioning in terms of delay and error-rate, consisting of {ground mobile users (MUs), indexed by $k$, GBSs, indexed by $g$, mmWave UAVs, indexed by $u$, and satellite, indexed by $s$.
We model the locations of GBSs as the independent homogeneous Poisson point process (HPPP), denoted by $\Phi_{\text{GBS}}$, with density $\lambda_{G}$, by implementing the stochastic geometry~\cite{10102783,5226957,6524460}. 
In addition, we model the 3D locations for UAVs as the 3D independent HPPP, denoted by $\Phi_{\text{UAV}}$, with intensity $\lambda_{U}$.


\subsection{Space-Air-Ground Integrated 3D Wireless Network Channel Model}

\subsubsection{3D Wireless Channel Model for the Satellite Network}

We assume that the wireless channel fading for the satellite link follows shadowed-Rician distribution~\cite{1198102}, which can efficiently describe the channel characteristics over wireless land mobile satellite communication systems. 
	{Correspondingly, we define the probability density function (PDF), denoted by $f_{|h_{k,s}|^{2}}(x)$, of the channel gain $|h_{k,s}|^{2}$ as follows~\cite{1198102}:
	\begin{equation}\label{equation_pdf}
		f_{|h_{k,s}|^{2}}(x)=\alpha_{s} e^{-\beta_{S} x}  \sideset{_1}{_{1}}{\mathop{F}}(\Gamma_{s},1,\delta_{s} x), \,\, x>0,
	\end{equation}
	where
	\begin{equation}\label{equation02}
		\begin{cases}
			\alpha_{s}=\frac{1}{2b_{s}}\left[\frac{2b_{s}\Gamma_{s}}{2b_{s}\Gamma_{s}+\Omega_{s}}\right]^{\Gamma_{s}};\\
			\beta_{S} = \frac{1}{2b_{s}};\\
			\delta_{s}= \frac{\Omega_{s}}{2b_{s}\left[2b_{s}\Gamma_{s}+\Omega_{s}\right]},
		\end{cases}
	\end{equation}
	where $\Omega_{s}$ denotes the average power of line-of-sight (LOS) component, $2b_{s}$ is the average power of the multipath component, $\Gamma_{s}\in[0,\infty]$ represents the Nakagami-$m$ parameter, and $_{1}F_{1}(\cdot,\cdot,\cdot)$ is the confluent hypergeometric function.
	For simplicity, we suppose that the Nakagami-$m$ parameter $\Gamma_{s}$ takes on integer values~\cite{6918460}.
	We can obtain
	\begin{align}
		&\sideset{_1}{_{1}}{\mathop{F}}(\Gamma_{s},1,\delta_{s} x)
		=e^{\delta_{s} x}\sum_{l=0}^{\Gamma_{s}-1}\frac{(-1)^{l}\left(1-\Gamma_{s}\right)_{l}\left[\delta_{s} x\right]^{l}}{(l!)^{2}}
	\end{align}
	where $(\cdot)_{l}$ denotes the Pochhammer symbol~\cite{LS2007}.
	Moreover, the cumulative probability function (CDF) of the channel gain $|h_{k,s}|^{2}$, denoted by $F_{|h_{k,s}|^{2}}(x)$, can be obtained as follows:
	\begin{equation}\label{equation02b}
		F_{|h_{k,s}|^{2}}(x)=\alpha_{s} \sum_{i=1}^{\infty}\frac{(\Gamma_{s})_{i}[\delta_{s}]^{i}}{(i!)^{2}[\beta_{S}]^{i+1}}\gamma(i+1, \beta_{S} x)
	\end{equation}
	where $\gamma(\cdot,\cdot)$ is the lower incomplete Gamma function.}

We consider that when there is no LOS link between the satellites and the ground mobile users, the mobile users choose to connect with the satellite network with a bias factor, denoted by $\phi_{S}$.
Then, the signal to interference and noise ratio (SINR), denoted by $\gamma_{k,s}$, when connecting with the satellite network is derived as follows:
\begin{equation}\label{equation04}
\gamma_{k,s}=\frac{\phi_{S}G_{S}{\cal P}_{s}|h_{k,s}|^{2}PL_{k,s}}{I_{k,s}+\sigma_{k,s}^{2}}
\end{equation}
where $h_{k,s}$ is the channel fading coefficient between the satellite and mobile user $k$, 
${\cal P}_{s}$ is the transmit power at the satellite, $G_{S}$ is the antenna gain at the satellite, $\sigma_{k,s}^{2}$ is the noise power in the satellite network, and $I_{k,s}$ is the aggregate interference power received from terrestrial interferers (nearby GBSs), defined as in the following equation: 
\begin{equation}
	I_{k,s}=\sum_{g\in\Phi_{\text{GBS}}}\phi_{S}{\cal P}_{g}|h_{k,g}|^{2}PL_{k,g}.
\end{equation}
where $PL_{k,s}$ and $PL_{k,g}$ are the link responses, which can be modeled as follows:
\begin{equation}\label{equation05}
	\begin{cases}
	PL_{k,s}\triangleq\left(\frac{c}{4\pi f_{s}}\right)^{2}[d_{k,s}]^{-\beta_{S}};\\
		PL_{k,g}\triangleq\left(\frac{c}{4\pi f_{g}}\right)^{2}[d_{k,g}]^{-\beta_{G}},
	\end{cases}
\end{equation}
where $c$ is the speed of light, $f_{s}$ and $f_{g}$ are the carrier frequencies for the communications between mobile user $k$ and the satellite and between mobile user $k$ and the GBS, $d_{k,s}$ and $d_{k,g}$ denote the distances between the satellite and mobile user $k$ and between the GBS and mobile user $k$, respectively, and $\beta_{S}$ and $\beta_{G}$ are the pathloss exponents over the satellite link and the ground link, respectively.

\subsubsection{3D Wireless Channel Model for the MmWave UAV Network}
Considering the wireless channel model for mmWave UAV link, the LOS connections between UAVs and terrestrial mobile users may be intermittently obstructed by terrestrial obstructions. 
The communication channel between ground-based mobile users and UAVs is conceptualized as a hybrid model, incorporating both LOS and non-line-of-sight (NLOS) pathways, factoring in their respective probabilities of occurrence. This framework allows for the calculation of pathloss gains for both LOS and NLOS transmissions~\cite{6863654}.
Accordingly, since the UAVs' 3D locations of are modeled by HPPP with density $\lambda_{U}$, the HPPP process can be further divided into two independent non-homogeneous PPPs, denoted by $\Phi^{\text{LOS}}_{\text{UAV}}$ with density $\lambda_{U}p_{k,u}^{\text{LOS}}(d_{k,u})$ and $\Phi^{\text{NLOS}}_{\text{UAV}}$, respectively, through applying independent thinning theorem~\cite{baccelli:inria-00403039}, where $p_{k,u}^{\text{LOS}}(d_{k,u})$ represents the LOS probability.
Thus, considering the distinct propagation characteristics of LOS and NLOS links, the pathloss gains, denoted by $PL_{k,u}^{\iota}$ with $\iota\in\{\text{LOS, NLOS}\}$, between the UAV and mobile user $k$ is given as follows:
\begin{align}
		PL_{k,u}^{\iota}=\xi_{k,u}^{\iota}\left(\frac{c}{4\pi f_{u}}\right)^{2}\left[d_{k,u}\right]^{-\beta_{U}^{\iota}}
\end{align}
where $f_{u}$ is the carrier frequency for the UAV in the mmWave UAV network, $\xi_{k,u}^{\iota}$ denotes the mean values of the excessive pathloss of LOS and NLOS links~\cite{8048502}, $\beta_{U}^{\iota}$ represents the pathloss exponent for LOS and NLOS links, and $d_{k,u}$ denotes the distance between mobile user $k$ and the UAV.
Define $z_{u}$ as the UAV's flight altitude, satisfying the minimum flight altitude $H_{\min}$ and the maximum flight altitude $H_{\max}$, respectively.
According to~\cite{6863654}, we can derive the LOS probability $p_{k,u}^{\text{LOS}}(d_{k,u})$ in the mmWave UAV network by implementing the elevation angle-dependent probabilistic LOS model as follows:
\begin{align}
	p_{k,u}^{\text{LOS}}(d_{k,u})=&\Bigg\{1+\nu^{\text{LOS}}_{1}\exp\Bigg\{-\nu^{\text{LOS}}_{2}\Bigg(\arctan\Bigg\{z_{u}
	\nonumber\\ 
	&\times\!\Bigg[\!\sqrt{(d_{k,u})^{2}\!-\!(z_{u})^{2}}\Bigg]^{-1}\Bigg\}\!-\!\nu^{\text{LOS}}_{1}\Bigg)\Bigg\}\Bigg\}^{-1}
\end{align}
where $\nu^{\text{LOS}}_{1}>0$ and $\nu^{\text{LOS}}_{2}>0$ denote the constants depending on the wireless SAGIN environment and $\arctan(\cdot)$ is the inverse tangent function.
Therefore, we can obtain total pathloss, denoted by $PL_{k,u}$, as follows:
\begin{equation}
	PL_{k,u}=	p_{k,u}^{\text{LOS}}(d_{k,u})PL_{k,u}^{\text{LOS}}+\left[1-p_{k,u}^{\text{LOS}}(d_{k,u})\right]PL_{k,u}^{\text{NLOS}}.
\end{equation}
Employing the sectored antenna model,  the directional antenna gain, expressed as $G_{U}$, specifically for the mmWave UAV network as $G_{U}=G^{\text{T}}_{U}G^{\text{R}}_{U}$, where the antenna gain for the mmWave UAVs is represented as $G^{\text{T}}_{U}$, while the mmWave antenna gain for the mobile users is signified as $G^{\text{R}}_{U}$.
We consider that when there exists the LOS link between the UAVs and the ground mobile users, the mobile users choose to connect with the UAV network with a bias factor, denoted by $\phi_{U}>0$.
Taking into account the 3D positioning of the UAVs, we can determine the mobile user' received power as follows:
\begin{align}\label{equation006}
	&P^{\text{R}}_{k,u}={\cal P}_{u}\phi_{U}G_{U}\left|h_{k,u}\right|^{2}PL_{k,u}
\end{align}
where ${\cal P}_{u}$ represents UAVs' transmit power and $h_{k,u}$ is the small-scale fading random variable, adhering to Nakagami-$m$ distribution characterized by integer parameter $\Gamma_{U}>0$.
Accordingly, the SINR, denoted by $\gamma^{}_{k,u}$, when connecting with mmWave UAV network, is derived as in the following equation:
\begin{align}\label{equation010}
	\gamma_{k,u}=&\Bigg\{{{\cal P}_{u}\phi_{U}G_{U}\left|h_{k,u}\right|^{2}\left(\frac{c}{4\pi f_{u} }\right)^{2}}\Bigg[p_{k,u}^{\text{LOS}}(d_{k,u})\xi_{k,u}^{\text{LOS}}
	\nonumber\\
	&\times \left[d_{k,u}\right]^{-\beta_{U}^{\text{LOS}}}	\!+\!\left[1\!-\!p_{k,u}^{\text{LOS}}(d_{k,u})\right]\xi_{k,u}^{\text{NLOS}}\left[d_{k,u}\right]^{-\beta_{U}^{\text{NLOS}}}\!\Bigg]\!\Bigg\}	\nonumber\\
	&\times \Bigg[{I_{k,u}+
		\left(\sigma_{k,u}\right)^{2}\Bigg]^{-1}
	}
\end{align}
where $\sigma_{k,u}$ represents the noise power while $I_{k,u}$ represents the aggregate interference power in the mmWave UAV network.
Owing to short wavelength characteristics, the implementation of directional beamforming within mmWave UAV-based wireless networks facilitates the achievement of \textit{interference isolation}. This technique effectively mitigates the influence of interference on adjacent networks, including those operating in both spatial and terrestrial domains.
Thus, we can determine the aggregate interference only considering the interference from other UAVs in the UAV network, which is given as follows:
\begin{align}\label{equation007}
	I_{k,u}&=\!\!\!\sum_{u'\in\Phi_{\text{UAV}}, \, u'\neq u}\!\!\!\!\!\!\!\!{{\cal P}_{u}\phi_{U}G_{U}\left|h_{k,u'}\right|^{2}\!\left(\frac{c}{4\pi f_{u} }\right)^{2}}\!\bigg[p_{k,u'}^{\text{LOS}}(d_{k,u'})\xi_{k,u'}^{\text{LOS}}
	\nonumber\\
	&\!\times\! \left[d_{k,u'}\right]^{-\beta_{U}^{\text{LOS}}}+\!\left[1\!-\!p_{k,u'}^{\text{LOS}}(d_{k,u'})\right]\xi_{k,u'}^{\text{NLOS}}
	\left[d_{k,u'}\right]^{-\beta_{U}^{\text{NLOS}}}\bigg].
\end{align} 
Accordingly, the aggregate interference power can be written by considering the interference powers over LOS and NLOS links, denoted by $I_{k,u}^{\iota}$ $(\iota\in\{\text{LOS, NLOS}\})$, respectively, as follows:
\begin{align}\label{equation033a}
	I_{k,u}=I_{k,u}^{\text{LOS}}+I_{k,u}^{\text{NLOS}}.
\end{align}
where
\begin{align}\label{equation033b}
		I_{k,u}^{\iota}= &\sum\limits_{u'\in\Phi^{\iota}_{\text{UAV}}, \, u'\neq u}{{\cal P}_{u}\phi_{U}G_{U}\left|h_{k,u'}\right|^{2}\left(\frac{c}{4\pi f_{u} }\right)^{2}}p_{k,u'}^{\iota} \nonumber\\
		&\qquad \qquad\qquad\times (d_{k,u'})\xi_{k,u'}^{\iota}\left[d_{k,u'}\right]^{-\beta_{U}^{\iota}}.
\end{align}

\subsection{The Channel Coding Rate Model Through Using FBC}

\subsubsection{The Normal Approximation}
Shannon's second theorem traditionally requires that the blocklength tends to infinity in order to accurately approximate the maximum coding rate. 
Nonetheless, when confronted with the constraints such as the rigorous demands for \textit{stringent delay and error-rate bounded QoS} for supporting mURLLC in contexts of finite blocklength, Shannon's capacity formula falls short of applicability.
Towards this end, we adopt the normal approximation method to characterize the maximum coding rate model while guaranteeing \textit{statistical} diverse QoS demands of mURLLC services when implementing FBC with a non-diminishing error-rate probability. 

\textit{Definition 1:} The decoding error probability, denoted by $\epsilon_{k,i}$, with coding blocklength, denoted by $n$, between mobile user $k$ and network $i$ $(i\in\{s,u\})$ for the FBC-based performance modeling scheme is defined  as follows~\cite{Yp2011}:
\begin{equation}\label{equation0018}
	\epsilon_{k,i}\approx \mathbb{E}_{\gamma_{k,i}}\left[ {\cal Q}\left(\frac{C_{k,i}-R^{*}_{k,i}}{\sqrt{V_{k,i}/n}}\right)\right]
\end{equation}
where  $\mathbb{E}_{\gamma_{k,i}}[\cdot]$ is the expectation taken with respect to the SINR $\gamma_{k,i}$, ${\cal Q}(\cdot)$ is the \textit{$Q$}-function, $R^{*}_{k,i}$ is the maximum coding rate between mobile user $k$ and network $i$ in the finite blocklength regime, and $C_{k,i}$ and $V_{k,i}$ are the {channel capacity}  and {channel dispersion} functions, which are given as follows:
\begin{equation}
	\begin{cases}	\vspace{1pt}
		C_{k,i}=\log_{2}\left(1+\gamma_{k,i}\right); \\
		V_{k,i}=1-\frac{1}{\left(1+\gamma_{k,i}\right)^{2}},
	\end{cases}
\end{equation}
where $\gamma_{k,i}$ is the SINR between in the satellite network in Eq.~\eqref{equation04} and the UAV network in Eq.~\eqref{equation010}, respectively.
Furthermore, by applying the normal approximation method, we can accurately approximate the maximum achievable coding rate $R^{*}_{k,i}$ by solving the above Eq.~\eqref{equation0018} for a given constrained decoding error probability in the finite blocklength regime.
Nonetheless, an exact closed-form solution for Eq.~\eqref{equation0018} remains elusive in the complex and dynamic environments of SAGINs. As an alternative approach, a precise approximation can be achieved by calculating the outage capacity function, as subsequently delineated.

\subsubsection{The Asymptotic Analysis}
\textit{Definition 2:} For a specified decoding error probability, an accurate approximation of the \textit{maximum achievable coding rate}, $R^{*}_{k,i}$ in bits per channel use  can be derived through the application of the asymptotic analysis method as follows~\cite{9941000}:
\begin{equation}\label{equation07}
	R^{*}_{k,i}=C^{\epsilon}_{k,i}+{\cal O}\left(\frac{\log n}{n}\right)
\end{equation} 
where $C^{\epsilon}_{k,i}$ represents the \textit{outage capacity} from network $i$ ($i\in\{s,u\}$) to mobile user $k$.

Therefore, it becomes essential to account for the outage probability to ascertain the outage capacity function. 
By setting the non-diminishing error probability to be $\epsilon_{k,i}$, the outage probability, denoted by $P^{\text{out}}_{k,i}$, for our developed performance modeling schemes is derived as in the following equation:
\begin{align}\label{equation032}
	P^{\text{out}}_{k,i}&
	=\text{Pr}\left\{\log_{2}\left(1+\gamma_{k,i}\right)<R^{*}_{k,i}\right\}.
\end{align} 
Accordingly, we can derive the outage capacity as follows~\cite{4137873}:
\begin{align}\label{equation033}
	C^{\epsilon}_{k,i}&=\sup\left\{R^{*}_{k,i}:P^{\text{out}}_{k,i}\leq\epsilon_{k,i} \right\}.
\end{align} 
By substituting Eq.~\eqref{equation032} into Eq.~\eqref{equation033}, we have 
\begin{align}\label{equation0114}
	C^{\epsilon}_{k,i}&=\sup\left\{R^{*}_{k,i}:\text{Pr}\left\{\log_{2}\left(1+\gamma_{k,i}\right)<R^{*}_{k,i}\right\}\leq\epsilon_{k,i} \right\}	\nonumber\\
	&=\sup\left\{R^{*}_{k,i}:\text{Pr}\left\{\gamma_{k,i}	<2^{R^{*}_{k,i}}-1\right\}\leq\epsilon_{k,i} \right\}.
\end{align} 
Previous works have shown that considering $F_{\gamma_{k,i}}(\cdot)$ as the complementary cumulative distribution function (CCDF) of the SINR $\gamma_{k,i}$, the outage capacity can be obtained by the following equation:
\begin{equation}\label{equation030}
	C^{\epsilon}_{k,i}=\log_{2}\left[1+F^{-1}_{\gamma_{k,i}}\left(1-\epsilon_{k,i}\right)\right]
\end{equation}
where $F^{-1}_{\gamma_{k,i}}(\cdot)$ is the inverse distribution function of $F_{\gamma_{k,i}}(\cdot)$.

	\section{The Decoding Error Probability Modeling in the Finite Blocklength Regime}\label{sec:DE}

	
\subsection{The Aggregate Interference Modeling}
The implementation of mmWave in UAV wireless networks introduces a scenario where the received signal may experience interference from nearby UAVs within the same network.
In addition, the GBSs can interfere with the satellite-MU communications in the satellite network. 
Thus, it becomes imperative to investigate the statistical characteristics of the aggregate interference across different networks, given its significant impact on calculating the decoding error probability over the SAGINs in the context of the finite blocklength domain. 
	
		\subsubsection{The Aggregate Interference in the Satellite Network}
The Laplace transform, denoted by ${\cal L}_{I_{k,s}}(s)$, with respect to the aggregate interference power $I_{k,s}$ between the satellite and mobile user $k$ is obtained as follows:
	\begin{align}\label{equation036}
		&{\cal L}_{I_{k,s}}(s)\!=\mathbb{E}_{I_{k,s}}\left[e^{-sI_{k,s}}\right]
		\nonumber\\
		&=\mathbb{E}_{\left|h_{k,s}\right|^{2}}\!\Bigg[\!\prod_{g\in\Phi_{\text{GBS}}}\!\!\!\exp\!\Bigg\{\!\!-s{\cal P}_{g}\phi_{S}
		\left|h_{k,g}\right|^{2}\!\left(\!\frac{c}{4\pi f_{g}}\!\right)^{2}\!\left[d_{k,g} \right]^{-\beta_{G}}\!\Bigg\}\!\Bigg]
		\nonumber\\
		&\,{\overset{(a)}{=}}
		\exp\!\Bigg\{\!-2\pi\lambda_{G}\!\!\int_{0}^{\infty}\!\!\Bigg[1\!-\!\mathbb{E}_{\left|h_{k,g}\right|^{2}}\!
		\Bigg[\exp\Bigg\{\!-s{\cal P}_{g}\phi_{S}\nonumber\\
		&\qquad\qquad\times\left|h_{k,g}\right|^{2}\!\left(\!\frac{c}{4\pi f_{g}}\!\right)^{2}\!\!\left[d_{k,g} \right]^{-\beta_{G}}\!\Bigg\}\Bigg]\Bigg]\!d_{k,g} dd_{k,g}\!\Bigg\}
	\end{align}	
	where {$(a)$ is obtained by applying the probability generating functional (PGFL) for HPPP $\Phi_{\text{GBS}}$,} ${\cal L}_{X}(s)$ denotes the Laplace transform with respect to a random variable $X\geq 0$, which is defined as follows:
	\begin{equation}
		{\cal L}_{X}(s)
		{\triangleq\int_{0}^{\infty}e^{-sx}f_{X}(x)dx}, \quad \text{ for  } s\geq 0
	\end{equation} 
where $f_{X}(x)$ denotes the probability density function in terms of the random variable $X$
	Given that the Nakagami-$m$ distribution in terms of the small-scale fading coefficient $h_{k,g}$ between the mobile users and GBSs, we can determine that $\left|h_{k,g}\right|^{2}$ is distributed according to the Gamma distribution, which is specified by its integer scale parameter $\Gamma_{G}$.
	Consequently, Eq.~\eqref{equation036} can be reformulated as presented in the subsequent equation:
	\begin{align}\label{equation035b}
		{\cal L}_{I_{k,s}}(s)=
		&\exp\Bigg\{-2\pi\lambda_{G}\int_{0}^{\infty}\Bigg[1-\Bigg\{1+s{\cal P}_{g}\phi_{S}\left(\frac{c}{4\pi f_{g}}\right)^{2}
		\nonumber\\
		&\qquad\,\,\,\times \left[d_{k,g} \right]^{-\beta_{G}}\Bigg\}^{-1}\Bigg]d_{k,g} dd_{k,g}\Bigg\}.
	\end{align}

	\subsubsection{The Aggregate Interference in the MmWave UAV Network}

Based on Eqs.~\eqref{equation033a} and~\eqref{equation033b}, the Laplace transform, denoted by ${\cal L}_{I_{k,u}}(s)$, with respect to the aggregate interference power $I_{k,u}$ over the mmWave UAV network level is determined as follows:
	\begin{align}
		{\cal L}_{I_{k,u}}(s)&={\cal L}_{I_{k,u}^{\text{LOS}}}(s){\cal L}_{I_{k,u}^{\text{NLOS}}}(s)
		=\mathbb{E}_{I_{k,u}}\left[e^{-s\left(I_{k,u}^{\text{LOS}}+I_{k,u}^{\text{NLOS}}\right)}\right]
	\end{align}
	where ${\cal L}_{I_{k,u}^{\text{LOS}}}(s)$ and ${\cal L}_{I_{k,u}^{\text{NLOS}}}(s)$ denote the Laplace transforms with respect to the aggregate interference powers across LOS and NLOS links, respectively, between the UAV and mobile user $k$, which can be obtained as follows:
	\begin{align}\label{equation027a}
	{\cal L}_{I_{k,u}^{\iota}}\!(s)\!=&\mathbb{E}_{\Phi^{\iota}_{\text{UAV}}}\Bigg[\!\mathbb{E}_{\left|h_{k,u'}\right|^{2}}\Bigg[\!\exp\Bigg\{\!-s\Bigg\{\sum\limits_{u'\in\Phi^{\iota}_{\text{UAV}}, \, u'\neq u}\!\!{\cal P}_{u}\phi_{U}\nonumber\\
			&\times\! G_{U}\left|h_{k,u'}\right|^{2}\left(\frac{c}{4\pi f_{u} }\right)^{2}p_{k,u'}^{\iota}(d_{k,u})\xi_{k,u'}^{\iota}\nonumber \\
			&\!\!\times\! \left[d_{k,u'}\right]^{-\beta_{U}^{\iota}}\!\Bigg\}\!\Bigg\}\!\Bigg]\Bigg]
			\nonumber \\
			\!=&\exp\!\bigg\{\!\!-2\pi\lambda_{U}\mathlarger{\int}_{0}^{\infty}\bigg\{1\!-\!\Big[1\!+\!s{{\cal P}_{u}\phi_{U}G_{U}\left(\frac{c}{4\pi f_{u} }\right)^{2}}\nonumber \\
			&\!\!\times\! p_{k,u'}^{\iota}(d_{k,u})\xi_{k,u'}^{\iota}\!\left[d_{k,u'}\right]^{-\beta_{U}^{\iota}}\!\!\Big]^{-1}\!\bigg\}
			p_{k,u'}^{\iota}(d_{k,u}) \nonumber \\
			&\!\!\times\! d_{k,u'} dd_{k,u'}\!\!\bigg\}.
	\end{align}

	\subsection{The Decoding Error Probability Modeling Using Normal Approximation}

	\subsubsection{The Decoding Error Probability in the Satellite Network}
We can obtain the closed-form expression of the decoding error probability function considering our proposed performance modeling schemes as in the subsequent theorem.

\begin{theorem}\label{theorem01}
	\textbf{If} the channel code defined by Definition 1 is applied to the proposed performance modeling schemes, \textbf{then} the decoding error probability $\epsilon_{k,s}$ for the satellite network is specified as follows:
	\begin{align}\label{theorem01_eq1}
		\epsilon_{k,s}\!\approx&\, \frac{\alpha_{s}}{\Gamma(k_{I_{k,s}})(\eta_{I_{k,s}})^{k_{I_{k,s}}}} \!\sum_{i=1}^{\infty}\!\frac{(\Gamma_{s})_{i}[\delta_{s}]^{i}}{(i!)^{2}[\beta_{S}]^{i+1}}\!\!
		\left[\frac{\beta_{S}\zeta^{\text{low}}_{k,s}}{\phi_{S}{\cal P}_{s}G_{S}PL_{k,s}}\!\right]^{i+1}
		\nonumber\\
		&\!\times\!\Gamma(i+1)\sum_{j=0}^{\infty}\frac{1}{\Gamma\left(i+j+2\right)}\left[\frac{\beta_{S}\zeta^{\text{low}}_{k,s}}{\phi_{S}{\cal P}_{s}G_{S}PL_{k,s}}\right]^{j}
		\nonumber\\
		&\!\times\!
		\left[\frac{\beta_{S}\zeta^{\text{low}}_{k,s}}{\phi_{S}{\cal P}_{s}G_{S}PL_{k,s}}+\frac{1}{\eta_{I_{k,s}}}\right]^{-(i+j+k_{I_{k,s}}+1)}\!\!\!\!\!\!\!\!\!\!\!\!\!\!\!\!\!\!\!\!\!\Gamma\left(i+j+k_{I_{k,s}}\!+1\right)
		\nonumber\\
		&\!+\!\Bigg[\frac{1}{2}+\vartheta_{k,s}\sqrt{n}\left(e^{R^{*}_{k,s}}-1\right)\Bigg]
		\frac{\alpha_{s}}{\Gamma(k_{I_{k,s}})(\eta_{I_{k,s}})^{k_{I_{k,s}}}}
		\nonumber\\
		&\!\times\! \sum_{i=1}^{\infty}\frac{(\Gamma_{s})_{i}[\delta_{s}]^{i}}{(i!)^{2}[\beta_{S}]^{i+1}}\Gamma(i+1)	\left[\frac{\beta_{S}}{\phi_{S}{\cal P}_{s}G_{S}PL_{k,s}}\right]^{i+1}
		\nonumber\\
		&\!\times\! \Bigg\{ \left(\zeta^{\text{up}}_{k,s}\right)^{i+1}\!\sum_{j=0}^{\infty}\frac{1}{\Gamma\left(i\!+\!j\!+\!2\right)}\!\left[\frac{\beta_{S}\zeta^{\text{up}}_{k,s}}{\phi_{S}{\cal P}_{s}G_{S}PL_{k,s}}\right]^{j}\!
		\nonumber\\
		&\!\times\!	\left[\frac{\beta_{S}\zeta^{\text{up}}_{k,s}}{\phi_{S}{\cal P}_{s}G_{S}PL_{k,s}}+\frac{1}{\eta_{I_{k,s}}}\right]^{-(i+j+k_{I_{k,s}}+1)}\!\!\!\!\!\!\!\!\!\!\!\!\!\!\!\!\!\!\!\!\!\Gamma\left(i\!+\!j\!+\!k_{I_{k,s}}\!+\!1\right)
		\nonumber\\
		&
		\!-\!\left(\zeta^{\text{low}}_{k,s}\right)^{i+1}\!\sum_{j=0}^{\infty}\frac{1}{\Gamma\left(i\!+\!j\!+\!2\right)}\left[\frac{\beta_{S}\zeta^{\text{low}}_{k,s}}{\phi_{S}{\cal P}_{s}G_{S}PL_{k,s}}\right]^{j}\!
		\nonumber\\
		&\times \!	\Bigg[\frac{\beta_{S}\zeta^{\text{low}}_{k,s}}{\phi_{S}{\cal P}_{s}G_{S}PL_{k,s}}
		+\frac{1}{\eta_{I_{k,s}}}\Bigg]^{-(i+j+k_{I_{k,s}}+1)}\!\!\!\!\!\!\!\!\!\!\!\!\!\!\!\!\!\!\!\!\!\Gamma\left(i\!+\!j\!+\!k_{I_{k,s}}\!+\!1\right)\!\!\Bigg\}\nonumber\\
		&-\!\vartheta_{k,s}\sqrt{n}\Lambda_{l}
	\end{align}
	where $\Gamma(\cdot)$ is the Gamma function, $k_{I_{k,s}}=\frac{\left(\mathbb{E}\left[I_{k,s}\right]\right)^{2}}{I\left[(I_{k,s})^{2}\right]}$,   $\eta_{I_{k,s}}=\frac{I\left[(I_{k,s})^{2}\right]}{\mathbb{E}\left[I_{k,s}\right]}$, where the mean and variance of $I_{k,s}$ are given, respectively, as follows:
	\begin{equation}\label{equation031a}
		\begin{cases}
			\!\mathbb{E}\left[I_{k,s}\right]\!=\!2\pi \lambda {\cal P}_{t} \sqrt{\frac{k_{\text{pd}+1}}{2k_{\text{pg}}}}\frac{R_{c}^{2-\alpha}}{2-\alpha};\\
			\!\text{Var}\left[I_{k,s}\right]\!=\!\pi\lambda [{\cal P}_{t}]^{2}k_{\text{pg}}(1+k_{\text{pg}})\eta_{\text{pg}}^{2}\frac{R_{c}^{2-2\alpha}}{1-\alpha},
		\end{cases}
	\end{equation}
	where $R_{c}$ is the radius of the operating region for the GBSs encircling the mobile user and
	\begin{equation}
		\begin{cases}
			\vartheta_{k,s}\triangleq\frac{1}{2\pi\sqrt{2^{2R^{*}_{k,s}-1}}};\\
			\zeta^{\text{low}}_{k,s}\triangleq2^{R^{*}_{k,s}-1} - \frac{1}{2\vartheta_{k,s}\sqrt{n}};\\
			\zeta^{\text{up}}_{k,s}\triangleq2^{R^{*}_{k,s}-1} + \frac{1}{2\vartheta_{k,s}\sqrt{n}},
		\end{cases}
	\end{equation}
	and
	\begin{align}\label{equation073b}
		&\Lambda_{l}=
		\frac{\alpha_{s}}{\Gamma(k_{I_{k,s}})(\eta_{I_{k,s}})^{k_{I_{k,s}}}}
		\sum_{j=0}^{\Gamma_{s}-1}\frac{(-1)^{j}\left(1-\Gamma_{s}\right)_{k,s}}{(j!)^{2}}
		\nonumber\\
		&\quad\!\!\!\!\!\!\times\!\!\left[ \frac{\delta_{s}}{\phi_{S}{\cal P}_{s}G_{S}PL_{k,s}}\right]^{j}(j+k_{I_{k,s}})!\frac{\eta_{I_{k,s}}^{j+k_{I_{k,s}}+1}}{j+2} \Bigg[
		\left(\zeta^{\text{up}}_{k,s}\right)^{j+2}
		\nonumber\\
		&\quad \!\!\!\!\!\!\times\!\!\! \sideset{_2}{_{1}}{\mathop{F}}\!\!\left(j\!+\!k_{I_{k,s}}\!+\!1,\!j\!+\!2,\!j\!+\!3,\!-\frac{\zeta^{\text{up}}_{k,s}\eta_{I_{k,s}}\left[\beta_{S}\!\!-\!\delta_{s}\!\right]}{\phi_{S}{\cal P}_{s}G_{S}PL_{k,s}}\right)
		\nonumber\\
		&\quad\!\!\!\!\!\!- \!\! \sideset{_2}{_{1}}{\mathop{F}}\!\!\left(\!j\!+\!k_{I_{k,s}}\!+\!1,j\!+\!2,j\!+\!3,-\frac{\eta_{I_{k,s}}\left[\beta_{S}-\delta_{s}\right]}{\phi_{S}{\cal P}_{s}G_{S}PL_{k,s}}\zeta^{\text{low}}_{k,s}\!\right)
		\nonumber\\
		&\quad\!\!\!\!\!\!\times  \!\left(\zeta^{\text{low}}_{k,s}\right)^{j+2}\Bigg].
	\end{align}
\end{theorem}

\begin{IEEEproof}
	The proof of Theorem~\ref{theorem01} is in Appendix A.
\end{IEEEproof}

\indent\textit{Remarks on Theorem~\ref{theorem01}:} Theorem~\ref{theorem01} provides an analysis of the decoding error probability for our developed modeling schemes.
Due to the complexity of the obtained closed-form expression for the decoding error probability, we opt for a more convenient approach by leveraging its asymptotic representation in the high signal-to-noise ratio (SNR) region in the following, simplifying the analysis to achieve a clearer understanding of the system's behavior while retaining its essential characteristics.

Considering high SNR region, i.e., ${\cal P}_{s}/\sigma_{k,s}\rightarrow\infty$, the asymptotic CDF of the channel fading gain $|h_{k,s}|^{2}$, denoted by $F^{\infty}_{|h_{k,s}|^{2}}(x)$, between the satellite and mobile user $k$ is derived as follows:
\begin{equation}\label{equation074}
	F^{\infty}_{|h_{k,s}|^{2}}\!(x)\!=\!\alpha_{s}\! \!\sum_{i=1}^{\infty}\!\frac{(\Gamma_{s}\!)_{i}[\delta_{s}]^{i}}{(i!)^{2}[\beta_{S}]^{i+1}}\frac{\left[\beta_{S} x\right]^{i+1}}{i+1}
	\!\approx\! \alpha_{s}x.
\end{equation}
Given the distribution of the interference power $I_{k,s}$, the asymptotic CDF of the SINR, denoted by $F^{\infty}_{\gamma_{k,s}}(x)$, between the satellite and the destination node can be derived as follows:
\begin{align}\label{equation075}
&	F^{\infty}_{\gamma_{k,s}}(x)
	=\int_{0}^{\infty}F^{\infty}_{|h_{k,s}|^{2}}\left(\frac{xy}{\phi_{S}{\cal P}_{s}G_{S}PL_{k,s}}\right)f_{I_{k,s}}(y)dy
	\nonumber\\
	&\quad	=\frac{\alpha_{s}x}{\phi_{S}{\cal P}_{s}G_{S}PL_{k,s}\Gamma(k_{I_{k,s}})(\eta_{I_{k,s}})^{k_{I_{k,s}}}}	\int_{0}^{\infty}y^{k_{I_{k,s}}}e^{-\frac{y}{\eta_{I_{k,s}}}}dy
	\nonumber\\
	&\quad=\frac{\eta_{I_{k,s}}^{k_{I_{k,s}}+1}\Gamma\left(k_{I_{k,s}}+1\right)\alpha_{s}x}
	{\Gamma(k_{I_{k,s}})(\eta_{I_{k,s}})^{k_{I_{k,s}}}\phi_{S}{\cal P}_{s}G_{S}PL_{k,s}}.
\end{align}
In the high SNR region, the asymptotic decoding error probability, denoted by $\epsilon^{\infty}_{k,s}$, is expressed as follows:
\begin{align}\label{equation076}
	\epsilon^{\infty}_{k,s}
		\approx&\,	F^{\infty}_{\gamma_{k,s}}\left(\zeta^{\text{low}}_{k,s}\right)\!+\!\Bigg[\frac{1}{2}\!+\!\vartheta_{k,s}\sqrt{n}\left(e^{R^{*}_{k,s}}-1\right)\Bigg]	\Bigg[F^{\infty}_{\gamma_{k,s}}\left(\zeta^{\text{up}}_{k,s}\right)
		\nonumber\\
		&-\!F^{\infty}_{\gamma_{k,s}}(\zeta^{\text{low}}_{k,s})\Bigg]-\!\vartheta_{k,s}\sqrt{n}\int_{\zeta^{\text{low}}_{k,s}}^{\zeta^{\text{up}}_{k,s}}xdF^{\infty}_{\gamma_{k,s}}(x)	\nonumber\\
	\!\approx&\,
	F^{\infty}_{\gamma_{k,s}}\left(\zeta^{\text{low}}_{k,s}\right)\!+\!\Bigg[\frac{1}{2}\!+\!\vartheta_{k,s}\sqrt{n}\left(e^{R^{*}_{k,s}}-1\right)\!\Bigg]\!\Bigg[F^{\infty}_{\gamma_{k,s}}\left(\zeta^{\text{up}}_{k,s}\right)\!
	\nonumber\\
	&\! -\!F^{\infty}_{\gamma_{k,s}}(\zeta^{\text{low}}_{k,s})\Bigg]\!-\!\vartheta_{k,s}\sqrt{n}\Bigg[\zeta^{\text{up}}_{k,s}F^{\infty}_{\gamma_{k,s}}(\zeta^{\text{up}}_{k,s})\!-\zeta^{\text{low}}_{k,s}
	\nonumber\\
	&\times F^{\infty}_{\gamma_{k,s}}(\zeta^{\text{low}}_{k,s})
\!-\int_{\zeta^{\text{low}}_{k,s}}^{\zeta^{\text{up}}_{k,s}}F^{\infty}_{\gamma_{k,s}}(x)dx\Bigg].
\end{align}
In addition, by substituting Eq.~\eqref{equation075} into Eq.~\eqref{equation076}, we have
\begin{align}\label{equation077}
	&\epsilon^{\infty}_{k,s}	\!\approx \frac{\eta_{I_{k,s}}^{k_{I_{k,s}}+1}\Gamma\left(k_{I_{k,s}}+1\right)\alpha_{s}\zeta^{\text{low}}_{k,s}}	{\Gamma(k_{I_{k,s}})(\eta_{I_{k,s}})^{k_{I_{k,s}}}\phi_{S}{\cal P}_{s}G_{S}PL_{k,s}}	\!+\!\Bigg[\frac{1}{2}\!+\!\vartheta_{k,s}\sqrt{n}
	\nonumber\\
	&\!\quad\!\times\! \left(e^{R^{*}_{k,s}}-1\right)\Bigg]	\frac{\eta_{I_{k,s}}^{k_{I_{k,s}}+1}\Gamma\left(k_{I_{k,s}}+1\right)\alpha_{s}}{\Gamma(k_{I_{k,s}})(\eta_{I_{k,s}})^{k_{I_{k,s}}}\phi_{S}{\cal P}_{s}G_{S}PL_{k,s}}
	\nonumber\\
	&\!\quad\!\times\! \left(\zeta^{\text{up}}_{k,s}\!-\!\zeta^{\text{low}}_{k,s}\right)\!-\!\vartheta_{k,s}\sqrt{n}	\frac{\eta_{I_{k,s}}^{k_{I_{k,s}}+1}\Gamma\left(k_{I_{k,s}}+1\right)\alpha_{s}}{\Gamma(k_{I_{k,s}})(\eta_{I_{k,s}})^{k_{I_{k,s}}}\phi_{S}{\cal P}_{s}G_{S}PL_{k,s}}
	\nonumber\\
	&\!\quad\!\times\!\!\Bigg[\!(\zeta^{\text{up}}_{k,s})^{2}\!-\!(\zeta^{\text{low}}_{k,s})^{2}	\!-\!\!\int_{\zeta^{\text{low}}_{k,s}}^{\zeta^{\text{up}}_{k,s}}\!\!\frac{\eta_{I_{k,s}}^{k_{I_{k,s}}\!+\!1}\Gamma\left(k_{I_{k,s}}+1\right)\alpha_{s}x dx}	{\Gamma(k_{I_{k,s}}\!)(\eta_{I_{k,s}}\!)^{k_{I_{k,s}}}\phi_{S}{\cal P}_{s}G_{S}PL_{k,s}}\!\Bigg]
	\nonumber\\
	&\,\,= \frac{\eta_{I_{k,s}}^{k_{I_{k,s}}+1}\Gamma\left(k_{I_{k,s}}+1\right)\alpha_{s}}	{\Gamma(k_{I_{k,s}})(\eta_{I_{k,s}})^{k_{I_{k,s}}}\phi_{S}{\cal P}_{s}G_{S}PL_{k,s}}	\Bigg\{\xi_{\text{low},s}\!+\!\Bigg[\frac{1}{2}\!+\!\vartheta_{k,s}\sqrt{n}
	\nonumber\\
	&\quad\!\times\!\left(\!e^{R^{*}_{k,s}}\!-\!1\right)\!\!\Bigg]\!\! \left(\zeta^{\text{up}}_{k,s}\!-\!\xi_{\text{low},s}\right)\!-\!\vartheta_{k,s}\sqrt{n}\left((\zeta^{\text{up}}_{k,s})^{2}\!-\!\xi_{\text{low},s}^{2}\!\right)^{\!2}
	\nonumber\\
	&\quad \!\times\!	\left[1-\frac{\eta_{I_{k,s}}^{k_{I_{k,s}}+1}\Gamma\left(k_{I_{k,s}}+1\right)\alpha_{s}}	{2\Gamma(k_{I_{k,s}})(\eta_{I_{k,s}})^{k_{I_{k,s}}}\phi_{S}{\cal P}_{s}G_{S}PL_{k,s}}\right]	\Bigg\}.
\end{align}

	\subsubsection{The Decoding Error Probability in the mmWave UAV Network}

We can derive the decoding error probability, denoted by $\epsilon_{k,u}$, for mmWave UAV link by considering channel code defined by Definition 2 for the proposed performance modeling schemes.
We approximately obtain the average decoding error probability through implementing Laplace transform with respect to the aggregate interference power for our proposed performance modeling schemes as follows: 
\begin{align}\label{equation040b}
	\epsilon_{k,u}
	&=\sum_{\ell=0}^{\Gamma_{U}}(-1)^{\ell} {\Gamma_{U} \choose \ell}	\mathbb{E}_{I_{k,u}}\Bigg[\exp\Bigg\{\!\!-\!\frac{\ell\eta_{U}\left[2^{R^{*}_{k,u}}-1\right]}{{\cal P}_{u}\phi_{U}PL_{k,u}}
	\nonumber\\
	&\quad \times \left[I_{k,u}+(\sigma_{k,u})^{2}\right] \Bigg\}\Bigg]
	\nonumber\\
	&=\sum_{\ell=0}^{\Gamma_{U}}(-1)^{\ell} {\Gamma_{U} \choose \ell}	\exp\left\{-\frac{\ell\eta_{U}\left[2^{R^{*}_{k,u}}-1\right]		(\sigma_{k,u})^{2}}{{\cal P}_{u}\phi_{U}PL_{k,u}}\right\}	\nonumber\\
	&\quad {\cal L}_{I_{k,u}}(s)\Bigg|_{s=\frac{\ell\eta_{U}\left[2^{R^{*}_{k,u}}-1\right]}{{\cal P}_{u}\phi_{U}PL_{k,u}}}.
\end{align} 
Then, by substituting  Eq.~\eqref{equation040b} back into Eq.~\eqref{equation027a}, we can obtain 
\begin{align}\label{equation051b}
	\epsilon_{k,u}
	&=\sum_{\ell=0}^{\Gamma_{U}}(-1)^{\ell} {\Gamma_{U} \choose \ell}	\exp\left\{-\frac{\ell\eta_{U}\left[2^{R^{*}_{k,u}}-1\right]		(\sigma_{k,u})^{2}}{{\cal P}_{u}\phi_{U}PL_{k,u}}\right\}				\nonumber\\
	& \!\!\!\times \!	{\cal L}_{I_{k,u}^{\text{LOS}}}(s)\Bigg|_{s=\frac{\ell\eta_{U}\left[2^{R^{*}_{k,u}}-1\right]}{{\cal P}_{u}\phi_{U}PL_{k,u}}}\!	{\cal L}_{I_{k,u}^{\text{NLOS}}}(s)\Bigg|_{s=\frac{\ell\eta_{U}\left[2^{R^{*}_{k,u}}-1\right]}{{\cal P}_{u}\phi_{U}PL_{k,u}}}.
\end{align} 
Considering high SNR region, i.e., ${\cal P}_{u}/\sigma_{u,s}\rightarrow\infty$, the asymptotic average decoding error probability, denoted by $\epsilon^{\infty}_{k,u}$, is derived as follows: 
\begin{align}\label{equation051c}
\epsilon^{\infty}_{k,u}
	=&\sum_{\ell=0}^{\Gamma_{U}}(-1)^{\ell} {\Gamma_{U} \choose \ell}	\exp\left\{-\frac{\ell\eta_{U}\left[2^{R^{*}_{k,u}}-1\right]}{{\cal P}_{u}\phi_{U}PL_{k,u}}\right\}				\nonumber\\
	& \!\times \!	{\cal L}_{I_{k,u}^{\text{LOS}}}(s)\Bigg|_{s=\frac{\ell\eta_{U}\left[2^{R^{*}_{k,u}}-1\right]}{{\cal P}_{u}\phi_{U}PL_{k,u}}}\!	{\cal L}_{I_{k,u}^{\text{NLOS}}}(s)\Bigg|_{s=\frac{\ell\eta_{U}\left[2^{R^{*}_{k,u}}-1\right]}{{\cal P}_{u}\phi_{U}PL_{k,u}}}.
\end{align} 

Within the context of our proposed SAGIN framework, the task of analytically characterizing the decoding error probability through the normal approximation method as outlined in Eq.~\eqref{equation0018} can rapidly evolve into a highly complex, or even infeasible, endeavor. 
This complexity suggests that obtaining an exact closed-form expression for the decoding error probability typically remains unattainable when applying FBC. Consequently, to overcome these analytical challenges, the outage probabilities for our proposed methodologies can be precisely approximated through the application of the Laplace transform, as will be elaborated in the subsequent section.

\subsection{The Outage Probability Modeling Using the Laplace Transform}
By using the asymptotic analysis method in Definition 2, we can derive the outage probability function in different network levels as follows.

	\subsubsection{The Outage Probability in the Satellite Network}

The outage probability, denoted by $P^{\text{out}}_{k,s}$, in the satellite network is derived as follows:
	\begin{align}
		P^{\text{out}}_{k,s}&=\text{Pr}\left\{\gamma_{k,s}<2^{R^{*}_{k,s}}-1\right\}
		\nonumber\\
		&=\text{Pr}\left\{\left|h_{k,s}\right|^{2}< \frac{\left[2^{R^{*}_{k,s}}-1\right]
			\left[I_{k,s}+(\sigma_{k,s})^{2}\right]}{{\cal P}_{s}\phi_{S}PL_{k,s}}\right\}
		\nonumber\\
		&=
		F_{\left|h_{k,s}\right|^{2}}\left(\frac{\left[2^{R^{*}_{k,s}}-1\right]
			\left[I_{k,s}+(\sigma_{k,s})^{2}\right]}{{\cal P}_{s}\phi_{S}PL_{k,s}}\right)
	\end{align} 
	where $F_{\left|h_{k,s}\right|^{2}}(\cdot)$ is the CDF in terms of $\left|h_{k,s}\right|^{2}$. 
Then, based on Eq.~\eqref{equation075}, we can derive the asymptotic average decoding error probability, denoted by $\overline{\epsilon}^{\infty}_{k,s}$, in the high SNR region as follows:
\begin{align}\label{equation051}
	\overline{\epsilon}^{\infty}_{k,s}
	&\!=\frac{\alpha_{s}\left[2^{R^{*}_{k,s}}-1\right]
		\left[I_{k,s}+(\sigma_{k,s})^{2}\right]}{{\cal P}_{s}\phi_{S}PL_{k,s}}.
\end{align}

	\subsubsection{The Decoding Error Probability in the mmWave UAV Network}
Considering the mmWave UAV network, the outage probability, denoted by $P^{\text{out}}_{k,u}$, is derived as in the following equation:
\begin{align}
	P^{\text{out}}_{k,u}&=\text{Pr}\left\{\gamma_{k,u}<2^{R^{*}_{k,u}}-1\right\}
	\nonumber\\
	&=
	F_{\left|h_{k,u}\right|^{2}}\left(\frac{\left[2^{R\left(\gamma_{k,u}\right)}-1\right]
		\left[I_{k,u}+(\sigma_{k,u})^{2}\right]}{{\cal P}_{u}\phi_{U}PL_{k,u}}\right)
\end{align} 
where $F_{\left|h_{k,u}\right|^{2}}(\cdot)$ represents the CDF with respect to $\left|h_{k,u}\right|^{2}$. 
Based on Eq.~\eqref{equation051},  the outage probability $P^{\text{out}}_{k,u}$ in the mmWave UAV network can be written through applying the results in~\cite{Incomplete_gamma_fun} as follows:
\begin{align}\label{equation040}
	P^{\text{out}}_{k,u}
	&\!=\!\left\{1\!-\!\exp\left\{-\frac{\eta_{U}\left[2^{R^{*}_{k,u}}\!-\!1\right]
		\left[I_{k,u}+(\sigma_{k,u})^{2}\right]}{{\cal P}_{u}\phi_{U}PL_{k,u}}\right\}\!\right\}^{\Gamma_{U}}
\end{align} 
	where 
\begin{equation}
	\eta_{U}\triangleq \Gamma_{U}(\Gamma_{U}!)^{-\frac{1}{\Gamma_{U}}}
\end{equation}
	where $(x!)$ represents the factorial operation.
Referencing the previous results in~\cite{yury2010}, the asymptotic outage probability is recognized as the saddle point approximation of the average decoding error probability in SAGINs. 
This approximation technique serves as a bridge, connecting the theoretical models of the decoding error probability in finite blocklength scenarios with their asymptotic counterparts, thereby offering a methodological framework for understanding and predicting the performance of communication systems as blocklength increases.

Note that the above equation offers an alternative by calculating the outage probability function, which in turn facilitates the determination of the decoding error probability. This approach yields practical and valuable mechanism for the modeling and accessing of the developed SAGINs, offering a feasible pathway to address the complexities involved.

\section{The $\epsilon$-Effective Capacity for Multi-QoS Framework in SAGINs}\label{sec:EC}

	\subsection{The Outage Capacity Modeling}


\subsubsection{The Outage Capacity in the Satellite Network}	
Based on Eq.~\eqref{equation074} in the high SNR region, we can derive the outage probability, denoted by $P^{\text{out},\infty}_{k,s}$, for our proposed schemes as follows:
\begin{align}\label{equation07401}
	P^{\text{out},\infty}_{k,s}
	&\!=\frac{\alpha_{s}\left[2^{R^{*}_{k,s}}-1\right]
		\left[I_{k,s}+(\sigma_{k,s})^{2}\right]}{{\cal P}_{s}\phi_{S}PL_{k,s}}.
\end{align} 
which is Eq.~\eqref{equation051}.
According to Eq.~\eqref{equation033}, we can establish the following relationship by equating the outage probability to $\overline{\epsilon}^{\infty}_{k,s}$ given by Eq.~\eqref{equation051}:
\begin{equation}\label{equation041}
	F_{\left|h_{k,s}\right|^{2}}\left(\frac{\left[2^{R^{*}_{k,s}}-1\right]
		\left[I_{k,s}+(\sigma_{k,s})^{2}\right]}{{\cal P}_{s}\phi_{S}PL_{k,s}}\right)=\overline{\epsilon}^{\infty}_{k,s}.
\end{equation}  
Thus, based on Eq.~\eqref{equation07401} and Eq.~\eqref{equation041}, we can determine the following relationship in the high SNR region:
\begin{equation}\label{equation042b}
	2^{R^{*}_{k,s}}-1=\frac{\overline{\epsilon}^{\infty}_{k,s}{\cal P}_{s}\phi_{S}PL_{k,s}}{\alpha_{s}I_{k,s}}.
\end{equation}
As a result, we can derive the outage capacity, denoted by $C^{\epsilon,\text{SA}}_{k,s}$, over the satellite network in the high SNR region as follows:
\begin{align}\label{equation0301}
	C^{\epsilon,\text{SA}}_{k,s}=&\log_{2}\left(1+\frac{\overline{\epsilon}^{\infty}_{k,s}{\cal P}_{s}\phi_{S}PL_{k,s}}{\alpha_{s}I_{k,s}}\right).
\end{align} 

\subsubsection{The Outage Capacity in the mmWave UAV Network} 
Similarly, we can establish the following relationship by equating the outage probability to $\epsilon_{k,u}$:
	\begin{equation}\label{equation048}
		F_{\left|h_{k,u}\right|^{2}}\left(\frac{\left[2^{R^{*}_{k,u}}-1\right]\left[I_{k,u}+(\sigma_{k,u})^{2}\right]}{{\cal P}_{u}\phi_{U}PL_{k,u}}\right)=\epsilon_{k,u}.
	\end{equation}
Accordingly to Eq.~\eqref{equation051}, we can obtain the following relationship:
\begin{equation}\label{equation042c}
	2^{R^{*}_{k,u}}-1=-\frac{{\cal P}_{u}\phi_{U}PL_{k,u}}{\eta_{U}	\left[I_{k,u}+(\sigma_{k,u})^{2}\right]}\log\left[1-\left(\epsilon_{k,u}\right)^{\frac{1}{\Gamma_{U}}}\right].
\end{equation}
As a result, the outage capacity, denoted by $C^{\epsilon,\text{UAV}}_{k,u}$, in the UAV network, is determined as follows:
\begin{align}\label{equation0053}
	C^{\epsilon,\text{UAV}}_{k,u}=&\log_{2}\Bigg\{1-\frac{{\cal P}_{u}\phi_{U}PL_{k,u}}{\eta_{U}	\left[I_{k,u}^{\text{LOS}}+I_{k,u}^{\text{NLOS}}+(\sigma_{k,u})^{2}\right]}
	\nonumber\\
	&\qquad\qquad\times \log\left[1-\left(\epsilon_{k,u}\right)^{\frac{1}{\Gamma_{U}}}\right]\Bigg\}.
\end{align} 

	\subsection{The $\epsilon$-Effective Capacity Function}
Traditionally, the concept of statistical QoS provisioning, particularly for delay-sensitive applications, has been extensively studied within the framework of queuing theory, where the focus is on characterizing queuing behavior under stochastic arrivals and service processes. 
In such systems, delay is considered as a key performance metric, and the goal is to describe the tail behavior of the delay distribution and ensure a certain delay violation probability.
Drawing upon the Large Deviations Principle (LDP), given adequate conditions, the queuing process, represented as $Q_{k,i}(l)$, tends toward a convergence in distribution towards a stochastic variable $Q_{k,i}(\infty)$, whereby the following condition is satisfied:
	\begin{equation}\label{equation0061}
		-\lim_{Q_{k,i}^{\text{th}}\rightarrow\infty}\frac{\log\left(\text{Pr}
			\left\{Q_{k,i}(\infty)>Q_{k,i}^{\text{th}}\right\}\right)}{Q_{k,i}^{\text{th}}}=\theta_{k,i}
	\end{equation}
	where $Q_{k,i}^{\text{th}}$ is the overflow threshold and $\theta_{k,i}$ $(i\in\{s,u\})$ is the \textit{QoS exponent} of queuing delay, which signifies the queuing delay's exponential decay rate of the delay violation probability. 
Eq.~(\ref{equation0061}) elucidates that the likelihood of the queuing process surpassing a specified overflow threshold diminishes exponentially at a rate determined by $\theta_{k,i}$.

Leveraging statistical QoS theory, it is feasible to approximate the delay-bound violation probability, denoted by $p^{\text{dv}}_{k,i}$, in relation to the maximum achievable coding rate. The random delay process, represented as $D_{k,i}$, can be quantitative measured as in the following equation:
\begin{align}\label{equation0062}
	p^{\text{dv}}_{k,i}=\text{Pr}\left\{D_{k,i}\geq D_{k,i}^{\text{th}}\right\}&\approx 
	\delta_{k,i} \exp\left\{-\theta_{k,i} R^{*}_{k,i}D_{k,i}^{\text{th}}\right\}
\end{align}
where $\delta_{k,i}$ is the probability that the queue is non-empty and $D_{k,i}^{\text{th}}$ is the delay constraint.
The QoS exponent plays a critical role in quantifying the delay-bound violation probability. Specifically, a smaller $\theta_{k,i}$ corresponds to a slower decay rate in the tail of the delay distribution, implying that the system only provides a looser QoS guarantee. 
In this case, the delay violation probability decreases more gradually as the delay-bound increases, indicating that the system is more likely to experience larger delays, making it suitable for applications with less stringent delay requirements.
Conversely, a larger $\theta_{k,i}$ leads to a faster decay rate in the delay-bound violation probability. In this scenario, the probability of exceeding a given delay-bound diminishes rapidly, which is essential for delay-sensitive applications.

	\subsubsection{The $\epsilon$-Effective Capacity in the Satellite Network}
With the advent of the next generation wireless communication networks, the challenges of statistical QoS provisioning have become significantly more complex. 
Wireless environments are characterized by variable and unpredictable conditions such as fading, interference, etc., which introduce additional stochasticity to the service processes. Moreover, new paradigms such as mURLLC place stringent requirements on delay and reliability, making it imperative to develop more advanced QoS models.

Therefore, we introduce a new concept of the \textit{$\epsilon$-effective capacity} taking into account both delay and reliability requirements in terms of statistical QoS provisioning.
	
\textit{Definition 3:} Considering the non-diminishing decoding error-probability, the concept of \textit{$\epsilon$-effective capacity} $EC_{k,s}^{\epsilon}\left(\theta_{k,s}\right)$ is introduced by representing the maximum constant arrival rate for a specific service process, under the stringent conditions of statistical delay and error-rate bounded QoS constraints within the finite blocklength domain, which is delineated as follows:
\begin{align}\label{equation017}
	EC_{k,s}^{\epsilon}\left(\theta_{k,s}\right)\triangleq&
	-\frac{1}{\theta_{k,s}}\log\Big\{\epsilon_{k,u} +\mathbb{E}\Big[\left(1-\epsilon_{k,u}\right)
	\nonumber\\
	&\qquad\qquad\qquad\times \exp\left\{-\theta_{k,s} n R^{*}_{k,u}\right\}\Big] \Big\}.
\end{align}
Considering the high SINR regime, we can obtain the asymptotic $\epsilon$-effective capacity, denoted by $EC_{k,s}^{\epsilon,\infty}\left(\theta_{k,s}\right)$, as follows:
\begin{align}
&EC_{k,s}^{\epsilon,\infty}\!\left(\theta_{k,s}\right)
		=\!
		-\frac{1}{\theta_{k,s}}\log\bigg\{\epsilon_{k,s} \!+\!\mathbb{E}_{I_{k,s}}\bigg[\left(1\!-\!\epsilon_{k,s}\right)
		\exp\!\bigg\{\!\!-\!\theta_{k,s}
			\nonumber\\
		&\qquad \times\! n  \log_{2}\left(1+\frac{\epsilon_{k,s}{\cal P}_{s}\phi_{S}PL_{k,s}}{\alpha_{s}I_{k,s}}\right)\bigg] \!\bigg\}
		\nonumber\\
		&\,\,=\!
		\!-\frac{1}{\theta_{k,s}}\!\log\!\Bigg\{\!\epsilon_{k,s} \!+\!\left(1\!-\!\epsilon_{k,s}\right)
		\mathbb{E}_{I_{k,s}}\!\!\Bigg[\!\!\left(\!1\!+\!\frac{\epsilon_{k,s}{\cal P}_{s}\phi_{S}PL_{k,s}}{\alpha_{s}I_{k,s}}\!\right)^{\!\!\!-\widetilde{\theta}_{k,s}}\!\!\Bigg] \!\Bigg\}
	\end{align}
where $\widetilde{\theta}_{k,s}\triangleq  {\theta_{k,s} n}/{ \log2}$.

\subsubsection{The $\epsilon$-Effective Capacity in the MmWave UAV Network}
The $\epsilon$-effective capacity, denoted by $EC_{k,u}^{\epsilon}\left(\theta_{k,u}\right)$, between mobile user $k$ and the UAV is derived as follows:
\begin{align}\label{equation060}
	EC_{k,u}^{\epsilon}\left(\theta_{k,u}\right)\!=&
	\!-\!\frac{1}{\theta_{k,u}}\!\log\!\vvast\{\!\epsilon_{k,u} \!+\!\left(1\!-\!\epsilon_{k,u}\right)	\mathbb{E}_{I_{k,u}}\!\Bigg[\!\bigg\{\!1-{\cal P}_{u}\phi_{U}
	\nonumber\\
	&\times \frac{PL_{k,u}}{\eta_{U}	\left[I_{k,u}^{\text{LOS}}+I_{k,u}^{\text{NLOS}}+(\sigma_{k,u})^{2}\right]}
	\nonumber\\
	&\times \log\left[1-\left(\epsilon_{k,u}\right)^{\frac{1}{\Gamma_{U}}}\right]\bigg\}^{-\widetilde{\theta}_{k,u}}\Bigg] \vvast\}.	
		\end{align}
Considering the high SINR regime, we can obtain the asymptotic $\epsilon$-effective capacity, denoted by $EC_{k,u}^{\epsilon,\infty}\left(\theta_{k,u}\right)$, as follows:
\begin{align}\label{equation061}
	&EC_{k,u}^{\epsilon,\infty}\!\left(\theta_{k,u}\right)
	\!=\!-\frac{1}{\theta_{k,u}}\log\!\Bigg\{\!\epsilon_{k,u} \!+\!\frac{1\!-\!\epsilon_{k,u}}{\left\{\!-\!\log\!\left[1\!-\!\left(\epsilon_{k,u}\right)^{\frac{1}{\Gamma_{U}}}\!\right]\!\right\}^{\widetilde{\theta}_{k,u}}}
	\nonumber\\
	&\qquad\times \left\{\frac{\eta_{U}}{{\cal P}_{u}\phi_{U}PL_{k,u}}
	\right\}^{\widetilde{\theta}_{k,u}}
	\mathbb{E}_{I_{k,u}}\left[\left[1+\frac{I_{k,u}}{(\sigma_{k,u})^{2}}\right]^{\widetilde{\theta}_{k,u}}\,\right]  \Bigg\}
	\nonumber\\
	&\,\,\,\overset{(b)}{=}-\!\frac{1}{\theta_{k,u}}\!\log\!\vvast\{\!\epsilon_{k,u} \!+\frac{1\!-\!\epsilon_{k,u}}{\left\{\!-\!\log\!\left[1\!-\!\left(\epsilon_{k,u}\right)^{\frac{1}{\Gamma_{U}}}\!\right]\!\right\}^{\widetilde{\theta}_{k,u}}}\nonumber\\
	&\!\times\! \left\{\frac{\eta_{U}}{{\cal P}_{u}\phi_{U}PL_{k,u}}
	\right\}^{\widetilde{\theta}_{k,u}}\!
	\!\sum_{\ell=0}^{\infty} {\widetilde{\theta}_{k,u} \choose \ell}\!	\mathbb{E}_{I_{k,u}}\!\!\left[\!\left(\frac{I_{k,u}^{\text{LOS}}\!+\!I_{k,u}^{\text{NLOS}}}{(\sigma_{k,u})^{2}}\right)^{\ell}\!\right]\!\!\vvast\}
\end{align}
where $(b)$ follows from binomial series.

\section{Performance Evaluations}\label{sec:results}
	
An extensive series of simulations have been undertaken to substantiate and assess the efficacy of our developed FBC-based strategies for enabling rigorous and diverse QoS requirements in the realm of 6G SAGINs. 
It is posited that the GBSs operate at a frequency of 2.4 GHz, whereas the mmWave UAVs function at 28 GHz. 
The simulation parameters include setting the UAV's maximum and minimum operational altitudes at $H_{\max}$ = 500 m and $H_{\min}$ = 10 m, respectively.
Furthermore, the pathloss exponent for the GBS network is determined as $\beta_{G}\in[3,4]$, while the pathloss exponents for LOS and NLOS links within the UAV network are defined as $\beta_{U}^{\iota}\in[2,4]$ $(\iota\in\{\text{LOS, NLOS}\})$ in the UAV network.

To validate and assess the effectiveness of the developed performance modeling schemes, we employ a user association mechanism for the SAGINs that is predicated on the criterion of maximum biased-received-power, as detailed in~\cite{9119462}. 
This particular scheme of user association is instrumental in deciding the network affiliation, i.e., the satellites or UAVs, for a given typical mobile user. 
The association decision is based on the selection of network $i$ with $i\in\{s,u\}$ that maximizes the biased-received-power metric, thereby optimizing the connectivity and service quality experienced by the user within the integrated network framework, i.e., we have 
\begin{equation}
	i = \arg\max_{j\in\{s,u\}}\left\{\frac{{\cal P}_{j}\phi_{j}G_{j}c^{2}}{\left(4\pi f_{j}\right)^{2}\left[d_{k,j}\right]^{\beta_{j}}}\right\}.
\end{equation}
Accordingly, with the bias factors set at $\phi_{U}=10$~dB and $\phi_{S}=0$ dB, Fig.~\ref{fig:02b} illustrates the user association probability of different networks against the UAVs' density for our developed performance modeling schemes.
It can be discerned that user association probability of the mmWave UAV link exhibits an increasing trend relative to the UAVs' density $\lambda_{U}$.
A larger user association probability of the mmWave UAV network indicates that more users choose to connect with the mmWave UAV network, leading to a reduced user association probability in the satellite network.

\begin{figure}[!t]
	\centering
	\includegraphics[scale=0.367]{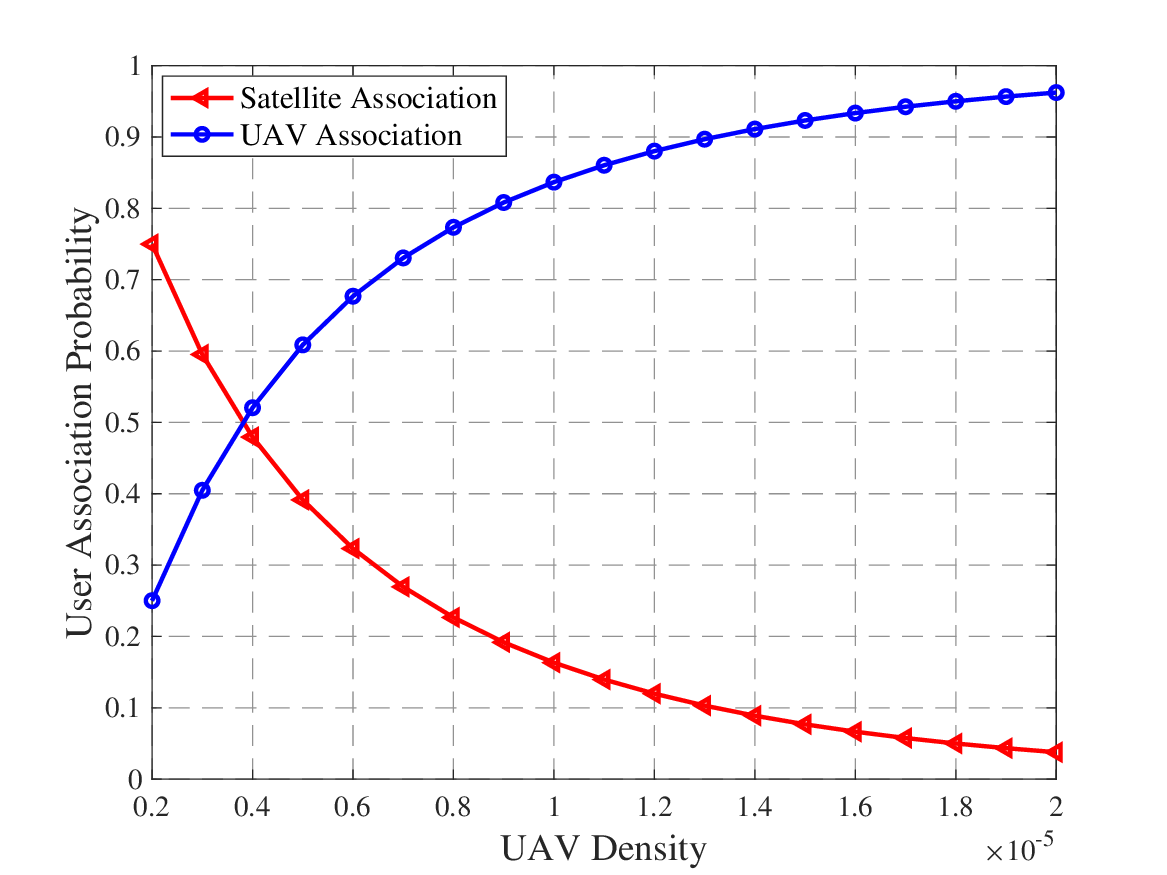}
	\caption{The user association probability vs. UAV density $\lambda_{U}$ for the developed performance modeling schemes.}
	\label{fig:02b}
\end{figure}

\begin{figure}[!t]
	\centering
	\includegraphics[scale=0.36]{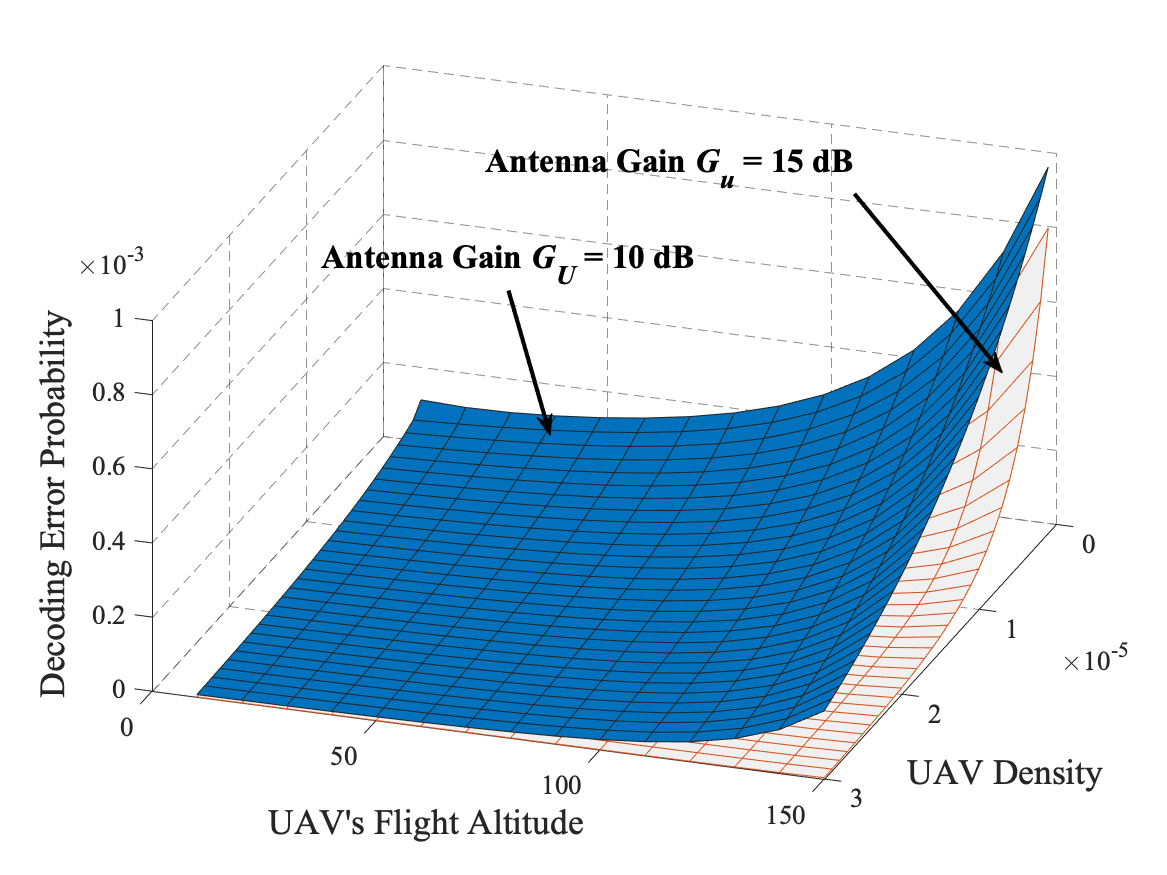}
	\caption{The decoding error probability  $\epsilon_{k,u}$ vs. UAV's flight altitude and UAV density $\lambda_{U}$  for the developed performance modeling schemes.}
	\label{fig:04}
\end{figure}

Figure~\ref{fig:04} presents the decoding error probability  $\epsilon_{k,u}$ plotted against the UAV density $\lambda_{U}$ and the UAV's flight altitude $z_{u}$ within UAV network considering various antenna gains.
Fig.~\ref{fig:04} reveals that decoding error probability inversely correlates with the UAV density, indicating a reduction in the decoding error probability as the number of UAVs increases..
Additionally, Fig.~\ref{fig:04} elucidates that a higher antenna gain facilitates a lower decoding error probability, indicating the significance of antenna gain in improving communication performance. 
Notably, the disparity between curves representing different antenna gains widens as the UAV's flight altitude increases, underscoring the impact of the flight altitude on the effectiveness of antenna gain in reducing decoding errors.

\begin{figure}[!t]
	\centering
	\includegraphics[scale=0.36]{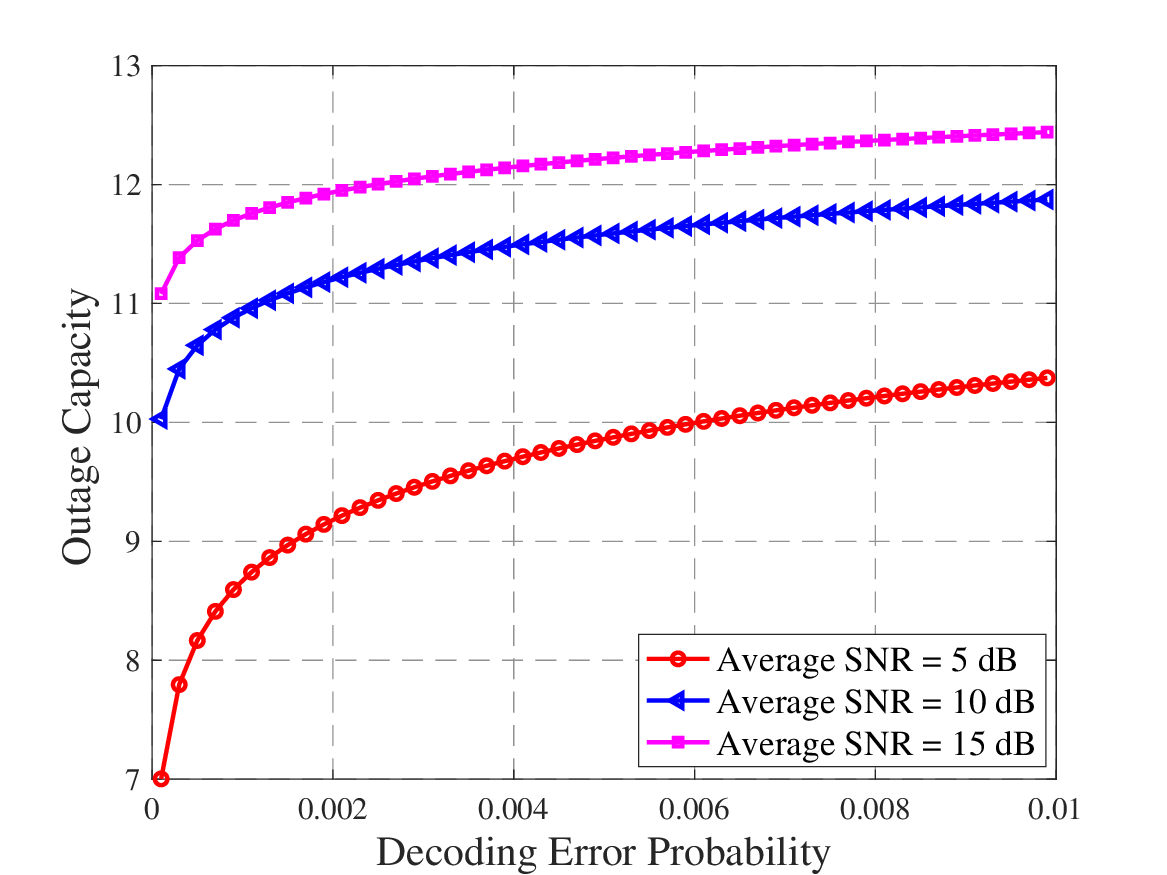}
	\caption{The outage capacity $P^{\text{out}}_{k,s}$ vs. decoding error probability $\epsilon_{k,s}$ in the satellite network for the developed performance modeling schemes.}
	\label{fig:05}
\end{figure}

With the bias factor set at $\phi_{S}=0$ dB and $\phi_{U}=10$~dB and setting the UAVs' density at $\lambda_{U}=15*10^{-6}$ BS/m$^2$, Fig.~\ref{fig:05} illustrates the outage capacity $P^{\text{out}}_{k,s}$ against the decoding error probability $\epsilon_{k,u}$ for the developed performance modeling schemes over the satellite network.
It is seen from Fig.~\ref{fig:05} that the outage capacity $P^{\text{out}}_{k,s}$ for satellites exhibits a positive correlation with the decoding error probability, indicating that larger decoding error probabilities are associated with increased outage capacities. 

\begin{figure}[!t]
	\centering
	\includegraphics[scale=0.36]{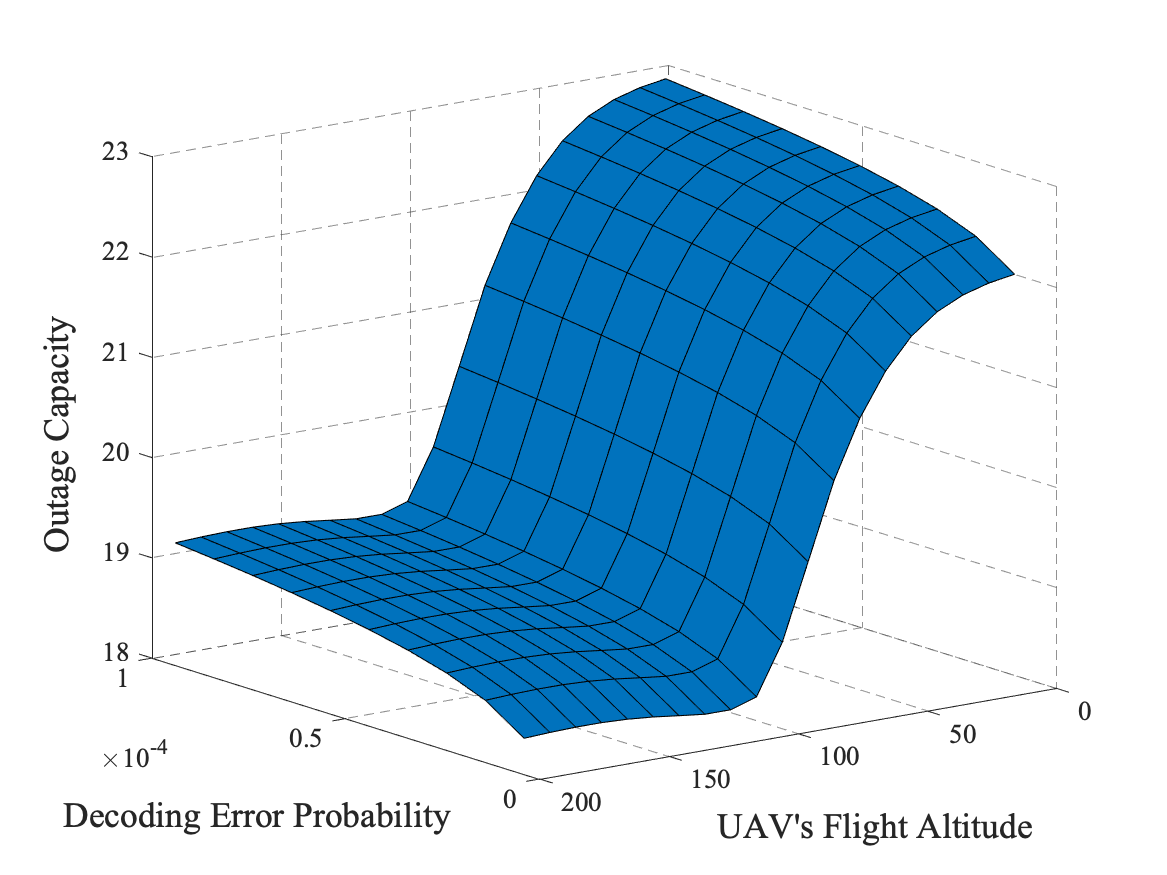}
	\caption{The outage capacity $P^{\text{out}}_{k,u}$ vs. UAV's flight altitude and decoding error probability for the developed performance modeling schemes.}
	\label{fig:06}
\end{figure}

Adjusting the bias factors at $\phi_{U}=\phi_{S}=0$ dB and setting the densities of the GBSs and UAVs $\lambda_{G}=\lambda_{U}=15*10^{-6}$ BS/m$^2$, Fig.~\ref{fig:06} delineates the outage capacity $P^{\text{out}}_{k,u}$ against decoding error probability $\epsilon_{k,u}$ and UAV's flight altitude $z_{u}$ in mmWave UAV network.
It is seen in Fig.~\ref{fig:06} that given a specified decoding error probability, the outage capacity inversely correlates with the UAV's flight altitude, diminishing as the altitude ascends and ultimately stabilizing at a particular value. This pattern underscores the impact of altitude on the operational efficiency of UAVs in terms of their capability to sustain communication reliability under specific decoding error probability.

\begin{figure}[!t]
	\centering
	\includegraphics[scale=0.36]{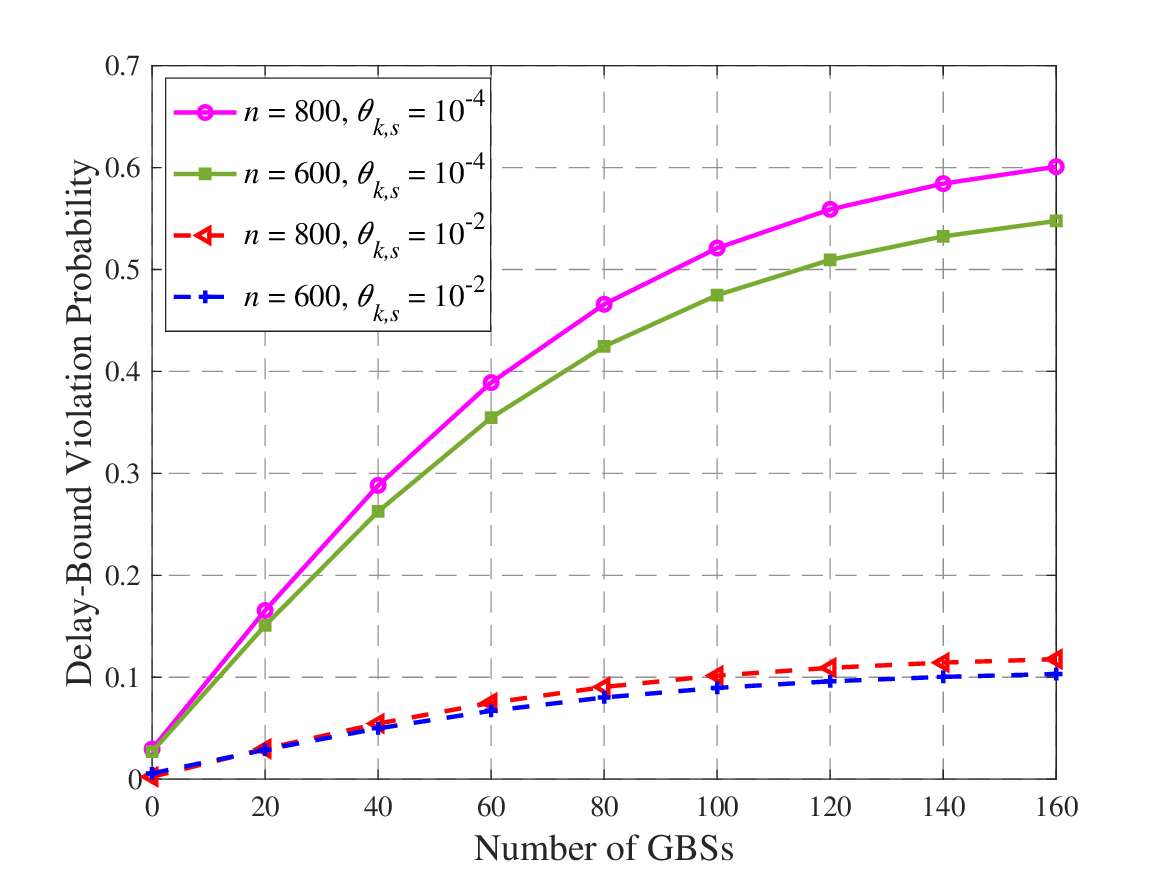}
	\caption{{The delay-bound violation probability vs. number of GBSs in the satellite network for the developed performance modeling schemes.}}
	\label{fig:010}
\end{figure}

Figure~\ref{fig:010} demonstrates the delay-bound violation probability $p^{\text{dv}}_{k,s}$ against the number of GBSs for our proposed performance modeling schemes with varying blocklength $n$. 
An analysis of Fig.~\ref{fig:010} reveals that the delay-bound violation probability exhibits a consistent decrease as a function of the blocklength, demonstrating a direct relationship between increased blocklength and enhanced reliability in terms of meeting delay constraints. 
Moreover, Fig.~\ref{fig:010} illustrates that imposing a larger number of GBSs leads to an increase in delay-bound violation probability, highlighting the impact of interference on the system's performance in adhering to delay-sensitive requirements.

\begin{figure}[!t]
	\centering
	\includegraphics[scale=0.36]{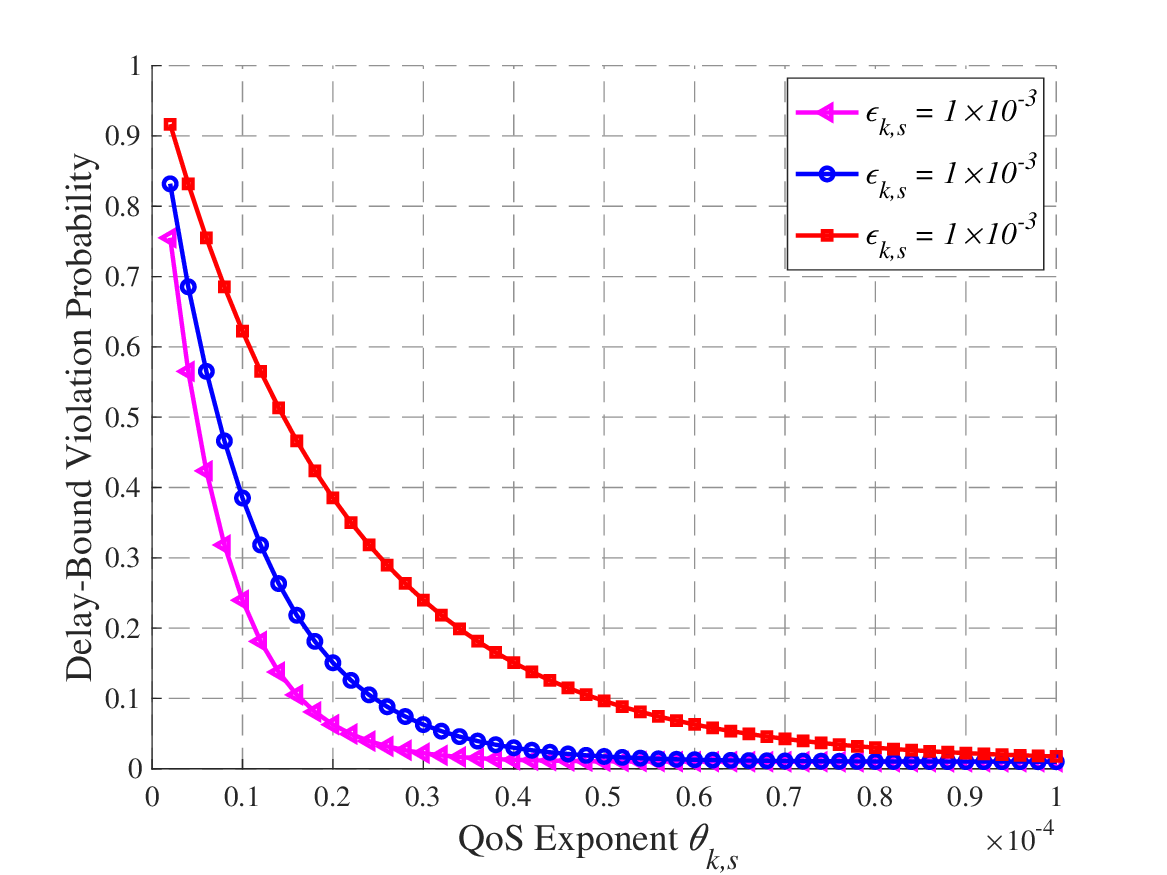}
	\caption{{The delay-bound violation probability vs. QoS exponent $\theta_{k,s}$ in the satellite network for the developed schemes.}}
	\label{fig:011}
\end{figure}

Figure~\ref{fig:011} illustrates the delay-bound violation probability $p^{\text{dv}}_{k,s}$ against the QoS exponent in the satellite network level. 
The analysis of Fig.~\ref{fig:011} indicates that the delay-bound violation probability consistently decreases as a function of the QoS exponent $\theta_{k,s}$. This trend signifies that a smaller QoS exponent delineates an upper limit on the delay-bound violation probability, while a larger QoS exponent establishes a lower threshold. Essentially, this relationship underscores the critical role of the QoS exponent $\theta_{k,s}$ in defining the system's capacity to adhere to delay constraints, where modifying $\theta_{k,s}$ can significantly influence the probability of delay-bound violations.

\begin{figure}[!t]
	\centering
	\includegraphics[scale=0.36]{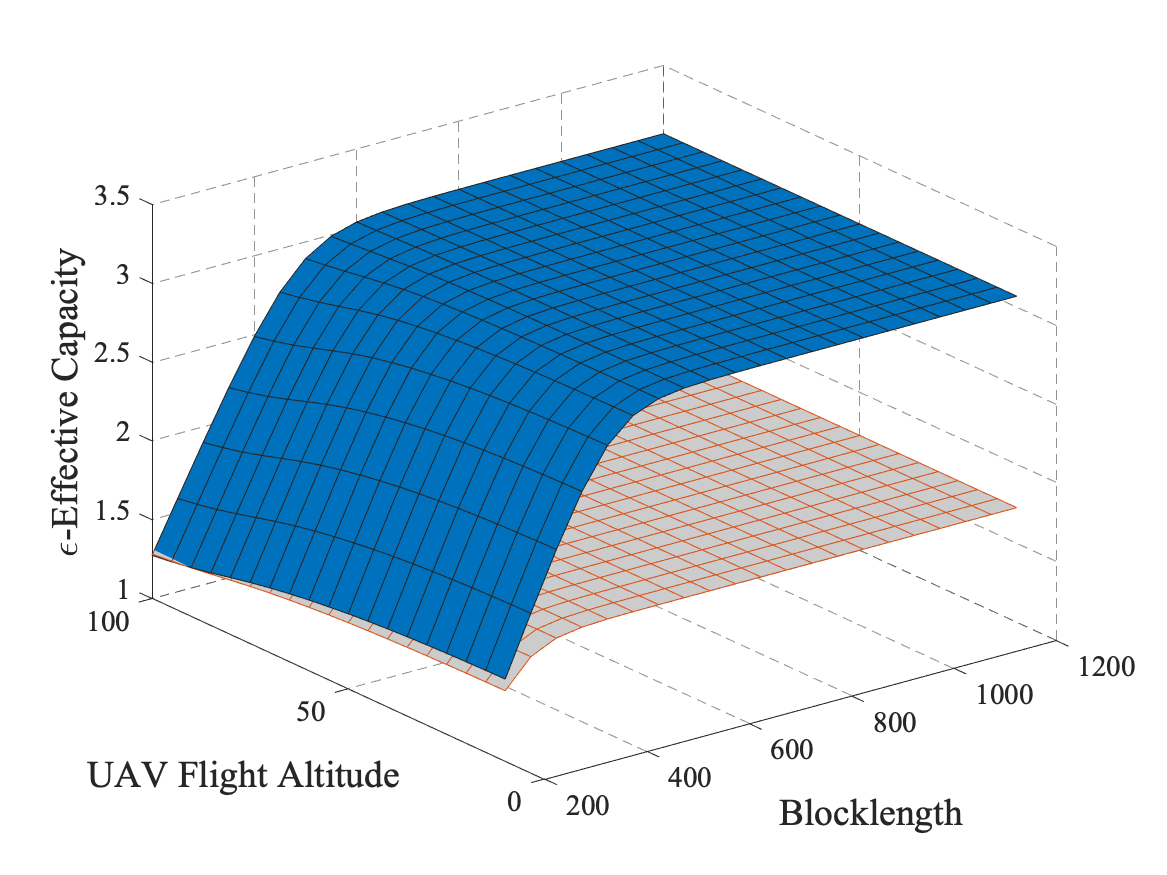}
	\caption{The $\epsilon$-effective capacity vs. UAV's flight altitude and blocklength in the mmWave UAV network for the developed performance modeling schemes.}
	\label{fig:08}
\end{figure}

With the bias factors setting at $\phi_{U}=\phi_{S}=0$ dB and the decoding error probability setting at $\epsilon_{k,g}=\{10^{-4},10^{-3}\}$, Fig.~\ref{fig:08} illustrates the $\epsilon$-effective capacity against the blocklength $n$ and UAV's flight altitude $z_{u}$ in mmWave UAV network considering varying decoding error probabilities, i.e., $\epsilon_{k,u}\in\{1\times 10^{-4},1\times 10^{-3}\}$.
The analysis of Fig.~\ref{fig:08} reveals that, for a given decoding error probability, the $\epsilon$-effective capacity experiences an increase with the blocklength, approaching a specific maximum value eventually.
Fig.~\ref{fig:08} shows that the $\epsilon$-effective capacity in mmWave UAV network exhibits a monotonically decreasing trend as $\theta_{k,u}$ increases. 
Additionally, Fig.~\ref{fig:08} showcases that the $\epsilon$-effective capacity in the UAV network escalates with an increase in the decoding error probability, highlighting the interplay between decoding error probability, blocklength, and UAV altitude in shaping the $\epsilon$-effective capacity within the considered communication framework.

{With the QoS exponent set to $\theta_{k,s}\in\{0.01,0.001\}$, Fig.~\ref{fig:09} illustrates the $\epsilon$-effective capacity against $n$ for our developed SAGINs as compares with the schemes without implementing the UAVs. 
Fig.~\ref{fig:09} reveals that given a QoS exponent $\theta$ and a specified blocklength, the inclusion of mmWave UAV network nearly doubles the $\epsilon$-effective capacity compared to the configuration without the mmWave UAV.
This observation underscores the enhanced performance in $\epsilon$-effective capacity delivered by the scheme incorporating the mmWave UAV as compared to the alternative lacking mmWave UAV support.

\begin{figure}[!t]
	\centering
	\includegraphics[scale=0.36]{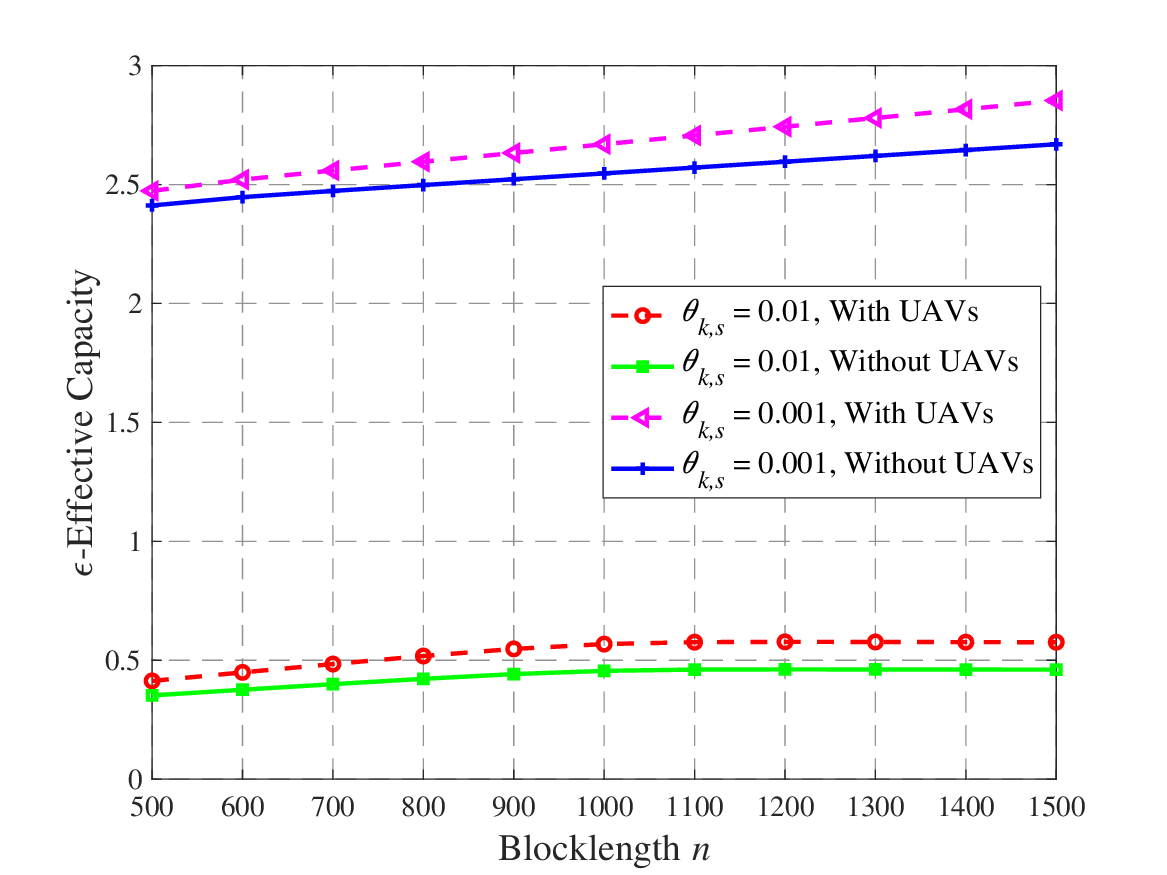}
	\caption{{The $\epsilon$-effective capacity vs. blocklength for the developed performance modeling schemes with and without implementing the mmWave UAV.}}
	\label{fig:09}
\end{figure}

	\section{Conclusions}\label{sec:conclusion} 

We have developed a set of analytical modeling frameworks specifically tailored to define and characterize the $\epsilon$-effective capacity function, thereby enabling the implementation of statistical delay and error-rate bounded QoS over SAGINs. 
In particular, we have established SAGIN system architectures alongside a channel coding model underpinned by FBC. Additionally, the decoding error probability model was derived utilizing the Laplace transform technique. 
Furthermore, we have designed modeling approaches for quantifying both the outage capacity and the $\epsilon$-effective capacity, crucial for the support of mURLLC within the confines of finite blocklength domain considering the high SNR region.
Finally, we perform a series of simulations to validate and assess the developed performance modeling schemes for mURLLC over SAGINs.

\begin{appendices}

	\section{Proof for Theorem 1}
We prove Theorem 1 through the following three steps, respectively.
	
	\underline{Step 1.}
	The decoding error probability $\epsilon_{k,s}$ in the satellite network is derived as follows:
	\begin{equation}\label{equation049}
		\epsilon_{k,s}\approx \int_{0}^{\infty} {\cal Q}\left(\frac{\sqrt{n}\left[C(x)-R^{*}_{k,s}\right]}{\sqrt{V(x)}}\right) f_{\gamma_{k,s}}\left(x\right) dx
	\end{equation}
	where $f_{\gamma_{k,s}}\left(x\right)$ represents the PDF of the SINR $\gamma_{k,s}$.
	Since $Q$-function has a complex form, it is difficult to find a closed-form expression for the decoding error probability. 
	Accordingly, we present an alternative approximation of the $Q$-function as follows:
	\begin{equation}\label{equation026}
		Q\left(\frac{C_{k,s}-R^{*}_{k,s}}{\sqrt{V_{k,s}/n}}\right)\approx \Psi(\gamma_{k,s})
	\end{equation}
	where $\Psi(\gamma_{k,s})$ is expressed as follows~\cite{BM2014}:
	\begin{equation}\label{equation027}
		\Psi\!(\gamma_{k,s}\!)\!=\!
		\begin{cases}
			\!1, \qquad\qquad\qquad\qquad\qquad\quad\,\,\,\, \gamma_{k,s}\!\leq\! \zeta^{\text{low}}_{k,s}; \\
			\!\frac{1}{2}\!-\!\vartheta_{k,s}\sqrt{n}\!\left(\!\gamma_{k,s}\!-\!2^{R^{*}_{k,s}-1}\!\right)\!, \,\,\,\,\,\, \zeta^{\text{low}}_{k,s}\!<\!\gamma_{k,s}\!<\!\zeta^{\text{up}}_{k,s}; \\
			\!0, \qquad\qquad\qquad\qquad\qquad\quad\,\,\,\, \gamma_{k,s}\!\geq\! \zeta^{\text{up}}_{k,s}.
		\end{cases}
	\end{equation}
	Taking expectation over Eqs.~(\ref{equation026}) and~(\ref{equation027}), we can obtain
	\begin{align}\label{equation028}
		\epsilon_{k,s}
		\!\approx&
		F_{\gamma_{k,s}\!}\left(\zeta^{\text{low}}_{k,s}\right)\!+\!\!\Bigg[\frac{1}{2}\!+\!\vartheta_{k,s}\sqrt{n}\left(e^{R^{*}_{k,s}}\!-\!1\!\right)\!\!\Bigg]
		\!\Bigg[\!F_{\gamma_{k,s}}\!\left(\zeta^{\text{up}}_{k,s}\right)
		\nonumber\\
		& 	-F_{\gamma_{k,s}}(\zeta^{\text{low}}_{k,s})\Bigg]
		\!-\vartheta_{k,s}\sqrt{n}\int_{\zeta^{\text{low}}_{k,s}}^{\zeta^{\text{up}}_{k,s}}xf_{\gamma_{k,s}}(x)dx
	\end{align}
	where $F_{\gamma_{k,s}}(x)$ is the CDF of the SINR $\gamma_{k,s}$.
	
	\underline{Step 2.} To analyze $\gamma_{k,s}$, we assume that the power gain between mobile user $k$ and the satellite decays exponentially with parameter $\widetilde{\alpha}$ and follows Gamma distributions with a shape parameter $k_{\text{pg}}$ and a scale parameter $\eta_{\text{pg}}$.	
To obtain a tractable model for the aggregate interference, we approximate the GBS interference distribution using the Gamma model considering Rayleigh fading. 
By Campbell's theorem, the mean aggregate interference is the same for all stationary point processes of the same intensity.
		We derive the characteristic function of aggregate interference, denoted by $\Phi_{I_{k,s}}$, as follows:
		\begin{equation}\label{equation071}
			\Phi_{I_{k,s}}(\omega)\!=\!\exp\!\left\{\!\!-2\pi\lambda_{G}\!\!\int\limits_{h_{k,s}}\!\!\int\limits_{\mathbb{R}}\!\left[1\!-\!e^{\jmath\omega x[d_{k,s}]^{-\widetilde{\alpha}}}\right]\!dh_{k,s}dd_{k,s}\!\right\}
		\end{equation}
		where $\jmath=\sqrt{-1}$.
		Based on Eq.~\eqref{equation071}, we can obtain the corresponding closed-form expression of the $i^{\text{th}}$ cumulant of $\Phi_{I_{k,s}}(\omega)$ as follows:
		\begin{equation}\label{equation072}
			\kappa_{I_{k,s}}(i)=\frac{1}{j^i}\frac{d^{i}}{d \omega^{i} }\frac{(\log  \Phi_{I_{k,s}}(\omega))}{1}\Big|_{\omega=0}
		\end{equation}
		Upon integrating Eq.~\eqref{equation071}, the result is obtained as follows~\cite{4453888}:
		\begin{equation}\label{equation078}
			\kappa_{I_{k,s}}(i)=\frac{2\pi\lambda_{G}}{i\widetilde{\alpha}-2}\mathbb{E}_{h_{k,s}}\left[\left[h_{k,s}\right]^{\frac{2}{\widetilde{\alpha}}}\right]
		\end{equation}
		Studies demonstrate that the aggregate received power from GBSs modeled by a HPPP can be effectively approximated using the Gamma distribution~\cite{6308772}.
		To obtain the closed form expressions of the PDF of aggregate interference power, denoted by $f_{I_{k,s}}(x)$, under the Gamma model, the PDF of aggregate interference can be approximately derived as follows:
		\begin{align}\label{equation77}
			f_{I_{k,s}}\!(x;k_{I_{k,s}},\eta_{I_{k,s}})
			&\!=\!\frac{x^{k_{I_{k,s}}\!-\!1}}{\Gamma(k_{I_{k,s}})(\eta_{I_{k,s}})^{k_{I_{k,s}}}}\!\exp\!\left\{\!-\frac{x}{\eta_{I_{k,s}}}\!\right\}\!.
		\end{align}
	Given the distribution of $I_{k,s}$, the CDF of the SINR $\gamma_{k,s}$ is given as follows:
	\begin{align}\label{equation053}
		F_{\gamma_{k,s}}(x)&=\text{Pr}\left\{\frac{\phi_{S}G_{S}{\cal P}_{s}|h_{k,s}|^{2}PL_{k,s}}{I_{k,s}+1} \leq x \Bigg|I_{k,s}\right\}
		\nonumber\\
		&=\int_{0}^{\infty}F_{|h_{k,s}|^{2}}\left(\frac{x(y+1)}{\phi_{S}{\cal P}_{s}G_{S}PL_{k,s}}\right)f_{I_{k,s}}(y)dy.
	\end{align}
	By assuming that the interference dominates the noise{, i.e., $I_{k,s}\gg1$, we can rewrite the SINR as follows:
		\begin{equation}\label{equation080}
			\gamma_{k,s}=\frac{\phi_{S}{\cal P}_{s}|h_{k,s}|^{2}G_{S}PL_{k,s}}{I_{k,s}}.
		\end{equation}
		Accordingly, we can rewrite Eq.~\eqref{equation053} as follows:
		\begin{align}\label{equation081}
			F_{\gamma_{k,s}}\!(x)
			=\int_{0}^{\infty}F_{|h_{k,s}|^{2}}\left(\frac{xy}{\phi_{S}{\cal P}_{s}G_{S}PL_{k,s}}\right)f_{I_{k,s}}(y)dy.
		\end{align}
		Then, plugging Eq.~\eqref{equation77} into Eq.~\eqref{equation081}, we have
		\begin{align}\label{equation055}
			&F_{\gamma_{k,s}}\!(x)\!
			=\frac{\alpha_{s}}{\Gamma(k_{I_{k,s}})(\eta_{I_{k,s}})^{k_{I_{k,s}}}} \sum_{i=1}^{\infty}\frac{(\Gamma_{s})_{i}[\delta_{s}]^{i}}{(i!)^{2}[\beta_{S}]^{i+1}}
			\nonumber\\
			&\quad\!\times\!\int_{0}^{\infty}\gamma\left(i\!+\!1,  \frac{\beta_{S}xy}{\phi_{S}{\cal P}_{s}G_{S}PL_{k,s}}\!\right) y^{k_{I_{k,s}}-1}e^{-\frac{y}{\eta_{I_{k,s}}}}dy  \nonumber\\
			&\,\,=\frac{\alpha_{s}}{\Gamma(k_{I_{k,s}})(\eta_{I_{k,s}})^{k_{I_{k,s}}}} \sum_{i=1}^{\infty}\frac{(\Gamma_{s})_{i}[\delta_{s}]^{i}}{(i!)^{2}[\beta_{S}]^{i+1}} \nonumber\\
			&\quad\times\!\left[\frac{\beta_{S}x}{\phi_{S}{\cal P}_{s}G_{S}PL_{k,s}}\right]^{i+1}\Gamma(i+1) \sum_{j=0}^{\infty}\frac{1}{\Gamma\left(i+j+2\right)} \nonumber\\
			&\quad\times\! \left[\!\frac{\beta_{S}x}{\phi_{S}{\cal P}_{s}G_{S}PL_{k,s}}\!\right]^{j} \!\!\!\!\int_{0}^{\infty}\!\!\!\! y^{i+j+k_{I_{k,s}}}e^{-\left[\frac{\beta_{S}x}{\phi_{S}{\cal P}_{s}G_{S}PL_{k,s}}+\frac{1}{\eta_{I_{k,s}}}\right]y}\!dy \nonumber\\
			&\,\,=\!\frac{\alpha_{s}}{\Gamma(k_{I_{k,s}})(\eta_{I_{k,s}})^{k_{I_{k,s}}}} \!\!\sum_{i=1}^{\infty}\!\!\frac{(\Gamma_{s}\!)_{i}[\delta_{s}]^{i}}{(i!)^{2}[\beta_{S}]^{i+1}}
			\!\!\left[\!\frac{\beta_{S}x}{\phi_{S}{\cal P}_{s}G_{S}PL_{k,s}}\!\right]^{\!i+1}
			\nonumber\\
			&\quad\!\!\times\!\Gamma(i+1)\sum_{j=0}^{\infty}\frac{1}{\Gamma\left(i+j+2\right)}\left[\frac{\beta_{S}x}{\phi_{S}{\cal P}_{s}G_{S}PL_{k,s}}\right]^{j}
			\nonumber\\
			&\quad\!\!\times\!
			\left[\frac{\beta_{S}x}{\phi_{S}{\cal P}_{s}G_{S}PL_{k,s}}\!+\!\frac{1}{\eta_{I_{k,s}}}\right]^{-(i+j+k_{I_{k,s}}+1)}
			\!\!\!\!\!\!\!\!\!\!\Gamma\left(i\!+\!j\!+\!k_{I_{k,s}}\!+\!1\right).
	\end{align}}
	Then, the PDF of the SINR $\gamma_{k,s}$ is derived as follows:
	\begin{align}\label{equation058}
		f_{\gamma_{k,s}}\left(x\right)&=\int_{0}^{\infty}(y+1)f_{|h_{k,s}|^{2}}\left(\frac{x(y+1)}{\phi_{S}{\cal P}_{s}G_{S}PL_{k,s}}\right)f_{I_{k,s}}(y)dy.
	\end{align}
	Similarly, by assuming that the interference dominates the noise, the PDF of the SINR $\gamma_{k,s}$ is rewritten as follows:
	\begin{align}\label{equation059}
		&f_{\gamma_{k,s}}\left(x\right)=\int_{0}^{\infty}yf_{|h_{k,s}|^{2}}\left(\frac{xy}{\phi_{S}{\cal P}_{s}G_{S}PL_{k,s}}\right)f_{I_{k,s}}(y)dy
		\nonumber\\
		&\,\,\,=	\frac{\alpha_{s}}{\Gamma(k_{I_{k,s}})(\eta_{I_{k,s}})^{k_{I_{k,s}}}}
		\sum_{l=0}^{\Gamma_{s}-1}\frac{(-1)^{l}\left(1-\Gamma_{s}\right)_{l}}{(l!)^{2}}
		\nonumber\\
		&\quad \times\!\!\left[\! \frac{\delta_{s}x}{\phi_{S}{\cal P}_{s}G_{S}PL_{k,s}}\!\right]^{l}\!\!
	\!\int_{0}^{\infty}\!\!\!\!
		y^{l+k_{I_{k,s}}}	e^{\!-\left\{\frac{\left[\beta_{S}-\delta_{s}\right] x}{\phi_{S}{\cal P}_{s}G_{S}PL_{k,s}}+\frac{1}{\eta_{I_{k,s}}}\right\}y}
		dy.
	\end{align}
	According to~\cite{LS2007}, we have
	\begin{align}\label{equation070}
		f_{\gamma_{k,s}}\left(x\right)&=	\frac{\alpha_{s}}{\Gamma(k_{I_{k,s}})(\eta_{I_{k,s}})^{k_{I_{k,s}}}}
		\sum_{l=0}^{\Gamma_{s}-1}\frac{(-1)^{l}\left(1-\Gamma_{s}\right)_{l}}{(l!)^{2}}
		\nonumber\\
		&\quad \times \left[ \frac{\delta_{s}x}{\phi_{S}{\cal P}_{s}G_{S}PL_{k,s}}\right]^{l}	(l+k_{I_{k,s}})!
		\nonumber\\
		&\quad \times
		\left\{\frac{\left[\beta_{S}-\delta_{s}\right] x}{\phi_{S}{\cal P}_{s}G_{S}PL_{k,s}}+\frac{1}{\eta_{I_{k,s}}}\right\}^{-l-k_{I_{k,s}}-1}.
	\end{align}
	To obtain the decoding error probability, the integral term in Eq.~\eqref{equation028} need to be obtained.
	Accordingly, we define the auxiliary function $\Lambda_{l}$ as follows:
	\begin{align}\label{equation073}
		\Lambda_{l}\triangleq&\int_{\zeta^{\text{low}}_{k,s}}^{\zeta^{\text{up}}_{k,s}}xf_{\gamma_{k,s}}(x)dx
		\nonumber\\
		=&
		\frac{\alpha_{s}}{\Gamma(k_{I_{k,s}})(\eta_{I_{k,s}})^{k_{I_{k,s}}}}
		\sum_{l=0}^{\Gamma_{s}-1}\frac{(-1)^{l}\left(1-\Gamma_{s}\right)_{l}}{(l!)^{2}}
		\nonumber\\
		& \times \left[ \frac{\delta_{s}}{\phi_{S}{\cal P}_{s}G_{S}PL_{k,s}}\right]^{l}	(l+k_{I_{k,s}})!\int_{\zeta^{\text{low}}_{k,s}}^{\zeta^{\text{up}}_{k,s}}x^{l+1}
		\nonumber\\
		& \times
		\left\{\frac{\left[\beta_{S}-\delta_{s}\right] x}{\phi_{S}{\cal P}_{s}G_{S}PL_{k,s}}+\frac{1}{\eta_{I_{k,s}}}\right\}^{-l-k_{I_{k,s}}-1}dx
	\end{align}
	which is Eq.~\eqref{equation073b}.
	By applying Eq.~(3.194) in~\cite{LS2007}, we can obtain Eq.~\eqref{equation073b}.
	
	\underline{Step 3.} By substituting Eqs.~\eqref{equation055} and~\eqref{equation073} into Eq.~\eqref{equation028}, the decoding error probability is obtained as specified by Eq.~\eqref{theorem01_eq1}. As a result, we can conclude the proof of Theorem~\ref{theorem01}.

\end{appendices}

	\nocite{*}
	\footnotesize
	\bibliographystyle{IEEEtran}
	\bibliography{myref.bib}
	
	\end{document}